\documentclass{IEEEtaes}

\ifCLASSINFOpdf
\else
\usepackage{graphicx}
\fi
\usepackage{url}

\usepackage{comment}

\usepackage{graphicx}
\usepackage{cite}
\usepackage{amsmath,amssymb,amsfonts}
\usepackage[linesnumbered,ruled]{algorithm2e}
\usepackage{textcomp}
\usepackage{amssymb}
\usepackage{amsmath}
\usepackage{graphicx}
\usepackage{graphics}
\usepackage{epstopdf}
\usepackage{caption}
\usepackage{xcolor,stfloats}
\usepackage{color}
\usepackage{float}
\usepackage{multicol}
\usepackage{array}
\usepackage{subfig}
\usepackage{bm}
\usepackage{cases}

\usepackage{siunitx}

\def\BibTeX{{\rm B\kern-.05em{\sc i\kern-.025em b}\kern-.08em
		T\kern-.1667em\lower.7ex\hbox{E}\kern-.125emX}}

\newlength\myindent
\newcommand{\blue}{\textcolor{black}}

\jvol{XX}
\jnum{XX}
\jmonth{XXXXX}
\paper{1234567}
\pubyear{2025}
\doiinfo{TAES.2025.Doi Number}

\begin{document}

	\title{Cognitive Non-Coherent Jamming Techniques for Frequency Selective Attacks}

	\author{Massimo Rosamilia}
	\member{Member, IEEE}
	\affil{Universita' degli Studi di Napoli ``Federico II'', DIETI, Via Claudio 21, I-80125 Napoli, Italy} 
	
	\author{Augusto Aubry}
	\member{Senior Member, IEEE}
	\affil{Universita' degli Studi di Napoli ``Federico II'', DIETI, Via Claudio 21, I-80125 Napoli, Italy} 
	
	\author{Vincenzo Carotenuto}
	\member{Senior Member, IEEE}
	\affil{Universita' degli Studi di Napoli ``Federico II'', DIETI, Via Claudio 21, I-80125 Napoli, Italy} 
	
	\author{Antonio De Maio}
	\member{Fellow, IEEE}
	\affil{Universita' degli Studi di Napoli ``Federico II'', DIETI, Via Claudio 21, I-80125 Napoli, Italy} 
	
	
	\receiveddate{Manuscript received XXXXX 00, 0000; revised XXXXX 00, 0000; accepted XXXXX 00, 0000.\\ The work of Massimo Rosamilia, Augusto Aubry, and Antonio De Maio was partially supported by the European Union under the Italian National Recovery and Resilience Plan (NRRP) of NextGenerationEU, partnership on ``Telecommunications of the Future'' (PE00000001 - program ``RESTART'').
		  }
	
	\corresp{{\itshape (Corresponding author: A. De Maio)}}
	
	\authoraddress{Massimo Rosamilia, Augusto Aubry, Vincenzo Carotenuto, and Antonio De Maio are with Universita' degli Studi di Napoli ``Federico II'', DIETI, Via Claudio 21, I-80125 Napoli, Italy, and also with the National Inter-University Consortium for Telecommunications, 43124 Parma, Italy (e-mail: massimo.rosamilia@unina.it; augusto.aubry@unina.it; vincenzo.carotenuto@unina.it; ademaio@unina.it).}
	
	\editor{
	}
	\supplementary{
		}

	\markboth{ROSAMILIA ET AL.}{Cognitive Non-Coherent Jamming Techniques for Frequency Selective Attacks}

	\maketitle

	\begin{abstract} This paper deals with the design of non-coherent jamming strategies capable of ensuring spectral compatibility with friendly radio frequency (RF) emitters. 
		The goal is achieved via a cognitive approach, which, after recognizing the presence of friendly RF systems within the bandwidth of interest (perception), synthesizes a jamming waveform (action) with spectral notches, that allows to interfere exclusively with opposite emissions. 	
		Two methods are proposed for the synthesis of the jamming signal. The former leverages optimization techniques for quadratically constrained quadratic problems (QCQP) where each constraint embeds the interference level tolerable by a specific friendly RF system. The latter is a very computationally efficient approach based on simple projections, allowing a control over the spectral notch positions and widths. At the analysis stage, the performance of the devised jamming techniques is firstly numerically analyzed in terms of spectral occupancy and autocorrelation characteristics. The impact of the quantization process involved in the digital-to-analog conversion (DAC) of the jamming waveforms is also examined, with a particular focus on the spectral shaping impairments resulting from reduced DAC resolution. Finally, waveform transmission and reception is experimentally assessed with software defined radio (SDR) devices.
	\end{abstract}
	
	\begin{IEEEkeywords} cognitive jamming, cognitive radio, ECM, ESM, waveform design
	\end{IEEEkeywords}
	
	\section{INTRODUCTION}
N{\scshape on-coherent} jamming is an electronic countermeasure (ECM) technique characterized by the transmission of an interfering signal whose frequency support is {comparable with or} much larger than the bandwidth of any {adversarial} radar {and/or} telecommunication system {within} the operational scenario. This allows to mask multiple radars/communications (and/or the entire agility bandwidth of a single system) simultaneously preventing data decoding {and target detection (possibly denying the accurate measurement of its features)} via {the transmission of} a noise-like non-coherent interference which can significantly deteriorate the actual signal-to-interference-plus-noise ratio (SINR)~\cite{stimson, adamy2001ew, pomr2}.
In general, depending on the specific application, a non-coherent jammer can operate in either a transmit-only noise generator mode or responsive noise transponder mode. The former can synthesize spot noise, barrage noise, blinking spot noise, and blinking barrage noise~{\cite{pomr2, Melvin2014-za, de2018introduction}}. The latter can provide responsive spot noise and noise cover pulse (noise-like signal introducing interference within all the range cells corresponding to its pulse width)~\cite[Part VIII]{stimson}.

Considering a congested and contested environment where both friendly and adversarial electromagnetic sources (radar and communication systems for instance) are present, the focus of the present study is {on} the design of {non-coherent} jamming waveforms capable of ensuring spectral coexistence with friendly systems (radars and/or communications) whose operating frequency support lies in the jamming activity bandwidth.  The goal is achieved resorting to a cognitive paradigm {by recognizing} the presence of non-adversarial {emitters} within the bandwidth of interest (perception stage) and subsequently synthesizing a jamming waveform interfering with {the} hostile radio frequency (RF) infrastructure {(action stage)}. {In particular, the perception stage envisaged by the cognitive framework can be implemented by leveraging} a {library}, {often} available at the jamming system, providing the parameters of friendly electromagnetic (EM) emissions~\cite{zhao2006radio, 4286322} and through an electronic support measurement (ESM) system that analyzes the bandwidth~{\cite{aubry2018multi, 9187976, 10891687}}, classifies emitters, extracts relevant spectral parameters, {and recognizes the presence of friendly spectral activities.}

The developed framework takes inspiration from the multitude of studies available in the open literature on the RF spectrum congestion problem which has attracted the interest of many scientists and engineers during the last decade and still represents one of the hot topics for the radar, communication, and signal processing {communities}~\cite{wicks2010spectrum,griffiths2014challenge, griffiths2014radar}. Notably, spectrum sharing strategies based on bespoke waveform designs have been envisioned as key enablers of radars and other RF systems {to cohabit} within the same frequency support~\cite{aubry2016optimization,bookBlunt, aubry2016forcing, aubry2020design, tang2018efficient, blunt2014polyphase, blunt2014polyphase2, alhujaili2020spectrally}. In this context, the cognitive radar (CR) paradigm has been recognized as a very promising solution to pursue efficient and smart usage of the frequency resources with limited cooperation among the overlaid active systems~\cite{farina2017impact,bookBlunt, guerci2020cognitive}. By leveraging the perception-action cycle (PAC), a CR acquires spectral awareness during the perception stage, and then, in the action phase, it dynamically selects probing waveforms to capitalize on the {available} spectral commodity~\cite{bookBlunt, farina2017impact,guerci2020cognitive,wicks2010spectrum,haykin2006cognitive,klemm2017novel, greco2018cognitive}. Perception involves the presence of radio environment maps storing data on the surrounding electromagnetic context as well as spectrum sensing modules whose goal is the real time update of the {emitter} databases~\cite{martone2015passive, martone2017spectrum,aubry2018multi,kovarskiy2020spectral}. Action demands the availability of waveform design algorithms optimizing the radar performance {while adhering to the spectral requirements} for cohabitation~\cite{he2010waveform, aubry2016optimization, aubry2016forcing,govoni2016enhancing,  huang2017radar, tang2018efficient, aubry2020design, 9693236}.

To accomplish the goal of synthesizing appropriate {non-coherent} jamming waveforms with controllable spectral notches (set according to the output of the perception stage), two methods are proposed.
The former {leverages} a quadratically constrained quadratic program (QCQP) formulation {aimed at controlling} the amount of interfering energy produced in the bandwidth occupied by friendly EM sources while maximizing the similarity with a noise-like signal.  {The latter} is a different approach enjoying fast deployment and {exploiting} simple projections, that only allows to control the spectral notches {width and} positions by nulling the components of the transmitted jamming signal in the stop-bands where the friendly emitters are operating.
{Moreover, the effect of quantization on the designed spectrally-shaped waveforms is also thoroughly examined. Notably,} at the transmission stage, in order to convert a digital waveform to an analog signal {to be} upconverted, amplified, and transmitted in the air via the antenna system, the digital-to-analog conversion (DAC) {demands} the waveform {being} {digitally quantized} to match its {suitable} resolution ({i.e., the number of bits used by the DAC}~\cite{van2013cmos}), in order to map each digital value to a specific voltage. This {quantization} process (related to the DAC resolution) can significantly affect the {spectral characteristics of the resulting analog waveform.} This effect {can become} critical for spectrally-shaped waveforms {developed} to {guarantee spectral coexistence in} a shared frequency interval, {e.g., in the context of integrated sensing and communication (ISAC)~\cite{7089157, bookBlunt, manzoni2024integrated}}. Indeed, as the number of DAC bits decreases, the resulting mismatch {(between synthesized and quantized waveform)} could {possibly impair} friendly transmitters, compromising effective electromagnetic spectrum sharing.

At the analysis stage, the performance of the devised strategies is numerically evaluated in terms of spectral {features} and autocorrelation characteristics of the synthesized signals in a scenario of practical interest. Moreover, the effect of the quantization process {on the devised spectrally-shaped waveforms} is investigated in terms of the spectral coexistence degradation induced by {a digital quantization with fewer number of bits}. As expected, the results show that lower DAC bit resolutions affect the notch depth, {impairing} the desired spectral cohabitation. The experimental validation of the devised strategies is then carried out using a specific hardware-in-the-loop testbed composed of some software defined radio (SDR) devices. The results show that the jamming waveforms designed via the proposed techniques comply with the spectral shape requirements and are hence suitable {candidates} for realizing {non-coherent} jamming while providing RF compatibility with friendly radiators.

{The remainder of the paper is organized as follows. Section II introduces the system model, detailing the spectral coexistence scenario and the proposed cognitive jammer architecture. In Section III, the problem of synthesizing non-coherent jamming waveforms for spectral coexistence is formulated and two cognitive design methods are presented; the former is based on QCQP while the latter resorts to simple orthogonal projections. Section IV discusses the impact of quantization on the spectral characteristics of the designed waveforms. In Section V, numerical simulations are performed to analyze the spectral and autocorrelation features of the non-coherent jamming waveforms synthesized with the proposed techniques. Section VI provides experimental validation using SDR devices to demonstrate the practical feasibility of the devised frameworks, whereas Section VII concludes the paper.}

\subsection{Notation}
We adopt the notation of using boldface for vectors $\bm{a}$ (lower case), and matrices $\bm{A}$ (upper case). The $n$th element of $\bm{a}$ and the $(m,l)$th entry of $\bm{A}$ are, respectively, denoted by $a(n)$ and $\bm{A}(m,n)$. The transpose and the conjugate transpose operators are denoted by the symbols $(\cdot)^\mathrm{T}$ and $(\cdot)^\dagger$, respectively. ${\mathbb{C}}^N$ is the set of $N$-dimensional vectors of complex numbers.  $\bm{I}$ denotes the identity matrix. The Euclidean norm of the vector $\bm{x}$ is denoted by $\Vert \bm{x}\Vert$. {For any $a \in \mathbb{R}$, $Q_b(a)$ denotes its value quantized using $b$ bits}. Finally, the letter $j$ represents the imaginary unit (i.e. $j=\sqrt{-1}$).

\section{SYSTEM MODEL}
\begin{figure*}[t]
	\centering
	\includegraphics[width=0.8\linewidth]{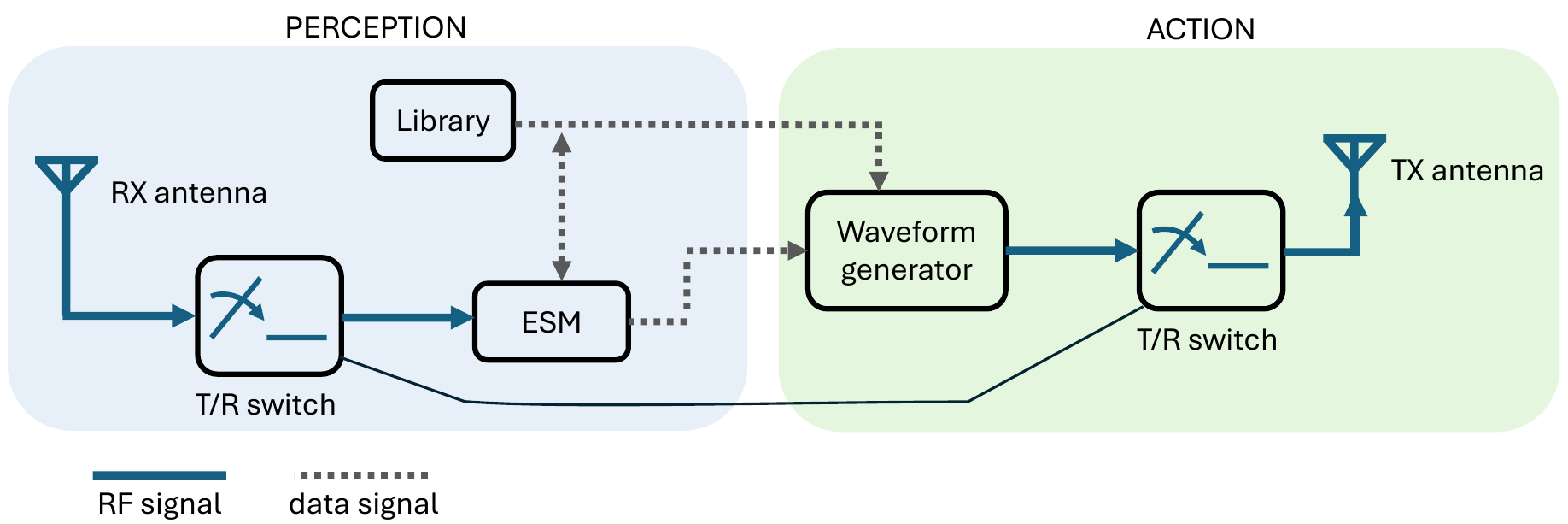}
	\caption{Conceptual architecture of a cognitive jammer equipped with an ESM system.}
	\label{fig:jammer_arch} 
\end{figure*}

Let us consider a cognitive jammer operating in the presence of possible {adversarial} {systems} as well as $K$ friendly {emitters}, each of them occupying a specific portion of the spectrum within the jammer frequency {operation range}.
At the base of the cognitive process there is the spectral awareness of the friendly EM emissions whose parameters are stored in a pre-canned library. Hence during the look-through phase, the jamming system {senses} the environment to understand which among the friendly emitters are active within the frequency interval object of the attack. This can be contextualized as a perception stage where T/R switch occasionally interrupts jamming transmission, and the ESM monitors the RF environment (Fig.~\ref{fig:jammer_arch}). 
Here the jammer gathers the information necessary to synthesize a {non-coherent} waveform capable of covering the desired frequency band while controlling its impact on {non-adversarial} overlaid transmissions. {Then, the action phase follows, where the appropriate waveform is synthesized (leveraging the information of the perception stage)} and is transmitted via the jammer antenna.

As to the RF {systems} coexisting with the jammer, it is supposed that each of them is operating over a frequency band $\Omega_k = [f_1^k, f_2^k], k=1,\dots, K$, where $f_1^k$ and $f_2^k$ denote the lower and upper normalized (with respect to the sampling frequency) frequencies for the $k$th system, respectively. 

{Let $c(t)$ be the baseband equivalent of the jammer transmitted waveform at the output of the arbitrary waveform generator whose complex input sequence (at the frequency rate equal to jammer bandwidth) is $c(i), \; i=1, \dots, N$.}
To ensure spectral compatibility with friendly emitters, the jammer has to properly shape {the} spectrum {of the transmitted signal} to manage the amount of interfering energy produced on the shared frequency bandwidths. {The average} jammer {signal} energy transmitted on the $k$th frequency band $\Omega_k$ is given by~\cite{aubry2016forcing}
\begin{equation}\label{eq:avg_energy}
	\frac{1}{f_2^k - f_1^k} \int_{f_1^k}^{f_2^k} S_c(f) df,
\end{equation}
where $S_c(f) = \left|\sum_{n=1}^{N} c(n) e^{-j 2 \pi f (n-1)}\right|^2 = N |\bm{c}^\dagger \bm{p}_f|^2$, with ${\bm{p}}_{{f}} = \frac{1}{\sqrt{N}} [1, \exp(-j2\pi {{f}}), \dots, \exp(-j2\pi(N-1) {{f}})]^\mathrm{T}$ the {temporal steering vector tuned to the} {normalized} frequency ${{f}}$.
That said, let us discretize the normalized frequency interval $[0,1]$ with steps $\Delta f=1/N$, wherein $f_i$ is $i$th element of the grid. Then, denoting by
\begin{equation}\label{eq:fourier_mat} 
	\bm{F}_N(m,n) = \frac{1}{\sqrt{N}} e^{-\frac{j 2 \pi (m-1)(n-1)}{N}}
\end{equation}
the $N \times N$ Fourier matrix scaled {by} $1/\sqrt{N}$, a {viable means} to quantify the average energy {level} transmitted on $\Omega_k$ is {through}
\begin{equation}
	\frac{N {\Delta f}}{f_2^k - f_1^k} \sum_{{f_i \in \Omega_k}} |\bm{c}^\dagger \bm{p}_{f_i}|^2 = \frac{1}{f_2^k - f_1^k} \sum_{{f_i \in \Omega_k}} |\bm{c}^\dagger \bm{p}_{f_i}|^2 = \bm{c}^\dagger{\bm{R}_N^k}\bm{c},
\end{equation}
where\footnote{{Unless otherwise specified, the subscript $N$ in $\bm{F}_N$, $\bm{Q}_N$, and $\bm{R}_N^k$ denotes the number of {their} rows and represents the number of points used to discretize {the frequency interval $[0,1]$}.}} ${\bm{R}_N^k} =\frac{1}{f_2^k - f_1^k} {\bm{Q}_N^k} {\bm{Q}_N^k}^\dagger$, with ${\bm{Q}_N^k}$ the submatrix of $\bm{F}_N$ whose columns correspond to the {discrete} frequencies belonging to $\Omega_k$.

Thus, {indicating} by $E_I^k, \; k=1,\ldots,K$, the acceptable level of disturbance\footnote{\blue{The values of $E_I^k$ can be obtained exploiting a-priori information that stems from the sensing process (as illustrated in~\cite{9109796}).}} on {$\Omega_k$}, which is related to the quality of service {demanded} by the $k$th {friendly emitter}, the transmitted {jamming} waveform has to comply with the constraints
\begin{equation}\label{eq:spectral_constr}
	\bm{c}^\dagger{\bm{R}_N^k}\bm{c}\leq E_I^k,\quad k=1,\ldots,K .
\end{equation}
By doing so, a detailed control of the interference energy produced on each shared frequency bandwidth is enforced.

\section{PROBLEM FORMULATION AND WAVEFORM DESIGN}
This section proposes two methods for designing jamming signals that aim {at hindering opponents (operating in the jammer bandwidth) sensing} capabilities by reducing {their} SINR without interfering with friendly emitters.
The former is based on a QCQP formulation {aimed at controlling} the amount of interfering energy produced in the frequency intervals occupied by {non-adversarial} EM sources while maximizing the similarity between the jamming waveform and a noise-like {reference} signal. The latter is a different approach with a reduced computational burden and based on simple projections. {However, it} only allows to control the spectral notches positions and width but not their depths. As pinpointed in the following, it is the limiting solution to the original design formulation.

\subsection{QCQP-based block waveform design}
In this design strategy, the similarity with a unit-norm noise-like signal $\bm{c}_0$  (characterized by a quasi-flat spectrum) is considered as figure of merit, in order to perform the spectral shaping {of} a {non-coherent} noise waveform. Precisely, considering the {waveform} composed of $L$ {disjoint} blocks of size $\bar{N}$, i.e.,
\begin{equation}
	\bm{c} = [\bm{c}_1^\mathrm{T}, \dots, \bm{c}_{L}^\mathrm{T}]^\mathrm{T},
\end{equation}
with\footnote{{Without loss of generality, it is assumed that $N = L \bar{N}$.}} $L =  N/\bar{N}$, the design approach {is tantamount to minimizing the distance $\|\bm{c}-\bm{c}_0 \|^2$ and} can be formulated as the following QCQP convex optimization problem
\begin{equation}\label{eq:qcqp_whole}
	\mathcal{P} \begin{cases}{\min\limits_{\bm{c}\in \mathbb{C}^{{N}}}} & \|\bm{c}-\bm{c}_0 \|^2\\ \mbox{s.t.} & 
			{\Vert {\bm{c}_l}\Vert ^2\leq 1/L}, \; l=1,\dots, L \\ &
		{\bm{c}}^{\dagger}{{\bm{R}_N^k}}{\bm{c}}\leq E_I^k,\,k=1,\ldots K\end{cases},
\end{equation}
whose optimal solution can be obtained in polynomial time with arbitrary precision.
{The norm constraint imposed on the individual blocks ${\bm{c}_i}$ of the sequence $\bm{c}$ is aimed at ensuring a bound on the transmitted energy, along with the avoidance of possible energy imbalances {of different segments in} the time domain.}

In practical scenarios where the waveform length $N$ is typically much larger than $10^{{6}}$, computing a solution to $\mathcal{P}$ involves handling a rather high computational and space (i.e., demanding the storage of large matrices) complexity. In such instances, this drawback can be {potentially} addressed by sequentially optimizing each block $\bm{c}_l, \; l=1, \dots, L$, leading to a (generally) sub-optimal, more efficient, and tractable design process.
Specifically, {by partitioning the reference code $\bm{c}_0$ in $L$ blocks of length $\bar{N}$, i.e., $\bm{c}_0 = [\bm{c}_{0,1}^T, \dots, \bm{c}_{0,l}^T, \dots, \bm{c}_{0,L}^{T}]^T$}, the design of the $l$th sub-block involves the maximization of the similarity with the reference {signal} ${\bm{c}_{0,l}}$ while satisfying tailored spectral constraints for the considered block, i.e.,
{\begin{equation}\label{eq:spectral_constr2}
	\bm{c}_l^\dagger \bm{R}_{\bar{N}}^k\bm{c}_l\leq E_I^k/L, \quad k=1,\ldots,K .
\end{equation}}

\blue{It is worth underlying that the duration of a single burst of the jamming waveform has a lower bound dictated by the sensing time plus the time necessary to synthesize the waveform. Moreover, it depends on the stationarity of the environment. In this respect, short segments allow for a more rapid adaptation to the environment with the deterioration of the noise-like behavior of the waveforms i.e., autocorrelation function and spectral behavior.}
\blue{Notice also that the approach of synthesizing blocks of the waveform, as opposed to the entire sequence, permits the transmission stage to start before the completion of the entire synthesis process. This approach reduces latency by enabling immediate transmission after the first block has been synthesized. Optimum balance is achieved when the synthesis time of a block is equal to the transmission time. In any case, the block synthesis time (in the case of a single processor employed at the design stage) must be less than the transmission time.}

Nevertheless, {synthesizing} each block {$\bm{c}_l$} independently leads to a lack of control {of} the spectral {behaviour} between consecutive blocks. {To guarantee a smooth transition between the subsequent blocks of samples and avoid {possible} spectral spuries, an heuristic but  effective solution technique is now illustrated. Specifically, the sequence $\bm{c}$ is progressively designed by employing a set of {partially overlapped} time windows (i.e., blocks) each including a segment of the signal already synthesized and a segment to be optimized.} Let {$\bm{\tilde{c}} \in \mathbb{C}^{\bar{N}}$} be a vector composed of the last {$W \in [0, \bar{N}/2]$} elements of the already designed block {concatenated} with the segment to optimize {at the current step} (of size $\bar{N}-W$). {The} optimization process then reformulates the quadratic constraints in terms of $\bm{\tilde{c}}$, thus ensuring a smooth spectral {behaviour}. Moreover, the total number of optimizations to perform (in addition to the one pertaining to the first block) is equal to $\tilde{L} = \left\lceil (N-\bar{N})/(\bar{N}-W) \right\rceil$.

\begin{algorithm}[t]
	\caption{QCQP block waveform design for selective {non-coherent jamming}}\label{alg:two}
	\KwIn{$N$, $\bar{N}$, $W$, $\{{\bm{R}_{\bar{N}}^k}\}_{k=1}^K$, $\{E_I^k\}_{k=1}^K$, {$\bm{c}_0$}}
	set $L = \left\lceil N/\bar{N} \right\rceil$ and $\tilde{L} = \left\lceil (N-\bar{N})/(\bar{N}-W) \right\rceil$\;
    {partition $\bm{c}_0$ in ${\tilde{L}+1}$ blocks as $\bm{c}_0 = [\bm{\bar{c}}_{0,1}^T, \bm{\bar{c}}_{0,2}^T, \dots, \bm{\bar{c}}_{0,\tilde{L}+1}^T]^T$, with $\bm{\bar{c}}_{0,1}$ of size $\bar{N}$ and ${\bm{\bar{c}}_{0,l}}$ of size $(\bar{N}-W), \, l=2, \dots, {\tilde{L}+1}$}\;
	compute $\bm{c}_1$ as a solution to~\eqref{eq:qcqp_first}\;
	\For{$l=[2, \dots, {\tilde{L}+1}]$}{
		compute $\bm{c}_l$ as a solution to~\eqref{eq:qcqp_block};
	}
	\KwOut{$\bm{c} = [\bm{c}_1^\mathrm{T}, \dots, \bm{c}_{{\tilde{L}+1}}^\mathrm{T}]^\mathrm{T}$}
\end{algorithm}

By leveraging the aforementioned guidelines and denoting by $\bm{c}_{l-1,W}$ the vector composed of the last $W$ elements of $\bm{c}_{l-1}$, the waveform design problem for the first and the $l$th ({with $l=[2, \dots, \tilde{L}+1]$}) block can be formulated as the following QCQP convex optimization problems
{\begin{equation}\label{eq:qcqp_first}
	\mathcal{P}_1 \begin{cases}{\min\limits_{\bm{\check{c}}_1\in \mathbb{C}^{\bar{N}}}} & \|\bm{\check{c}}_1-\bm{\bar{c}}_{0,1} \|^2\\ \mbox{s.t.} & 
		\Vert {\bm{\check{c}}_1}\Vert ^2\leq 1/L \\ &	
		{\bm{\check{c}}_1}^{\dagger}{{\bm{R}_{\bar{N}}^k}}{\bm{\check{c}}_1}\leq E_I^k / L,\,k=1,\ldots K\end{cases}
\end{equation}
and
\begin{equation}\label{eq:qcqp_block}
	\mathcal{P}_l \begin{cases}{\min\limits_{{\bm{\check{c}}_l}\in \mathbb{C}^{\bar{N}{-W}}}} & \|\bm{\check{c}}_l-{\bm{\bar{c}}_{0,l}} \|^2\\ \mbox{s.t.} & 
		\bm{\tilde{c}} = [{\bm{c}}_{l-1,W}^\mathrm{T}, {\bm{\check{c}}_l}^\mathrm{T}]^\mathrm{T} \\ &		\Vert {\bm{\tilde{c}}}\Vert ^2 \leq 1/{L}\\ &
		{\bm{\tilde{c}}}^{\dagger}{{\bm{R}_{\bar{N}}^k}}{\bm{\tilde{c}}}\leq E_I^k/{L},\,k=1,\ldots K\end{cases}
\end{equation}
respectively, where the reference code $\bm{c}_0$ is partitioned in ${\tilde{L}+1}$ blocks as $\bm{c}_0 = [\bm{\bar{c}}_{0,1}^T, \bm{\bar{c}}_{0,2}^T, \dots, \bm{\bar{c}}_{0,\tilde{L}+1}^T]^T$, with $\bm{\bar{c}}_{0,1}$ of size $\bar{N}$ and ${\bm{\bar{c}}_{0,l}}$ of size $(\bar{N}-W), \, l=2, \dots, {\tilde{L}+1}$.}
Remarkably,~\eqref{eq:qcqp_first} and~\eqref{eq:qcqp_block} can be {practically handled} being them computationally and space efficient.

{Summarizing,} by combining block-wise optimization with different reference {sequences} (corresponding to different portions of the original reference) and {tailored} spectral constraints, the proposed approach achieves an affordable computational {effort} while effectively mimicking the characteristics of white noise within the {enemy} bandwidth.
\textbf{Algorithm 1} {specifies} the proposed block-wise waveform design strategy\footnote{{A more efficient implementation (in terms of memory usage) of \textbf{Algorithm 1} involves the run-time generation of the reference {signal} $\bm{\bar{c}}_{0,l}$ when the $l$th block is optimized, instead of requiring the storage of the sequence $\bm{c}_0$.}} for selective {non-coherent} jamming.
\blue{It is also important to observe that while increasing the number of blocks (i.e., using smaller blocks) may offer advantages in terms of design speed, it comes at the cost of a reduced control over spectral shaping and notch depth (see Section~\ref{section:numerical_sim}-\ref{sub:incr_L}).}

\subsection{Spectral notching via orthogonal projection}
This section presents a {different approach} for spectral shaping, leveraging projections onto forbidden frequency bands. While lacking the detailed control offered by the QCQP approach, it provides rapid spectral notch placement.
As a matter of fact, it avoids solving optimization problems, making it particularly attractive for highly dynamic scenarios where the spectrum occupancy frequently changes in an unpredictab{le} way. 

The core idea behind this method lies in projecting the reference signal onto the subspace orthogonal to the steering vectors spanning {the} stop frequency bands. 
In a nutshell, being $\bm{x}\in \mathbb{C}^N$ a $N$-dimensional {digital} signal, it is possible to effectively removing from $\bm{x}$ the spectral components at {specific} frequencies $[f_1, \dots, f_S]$ by computing
\begin{equation}
	\bm{\bar{x}} = (\bm{I} - \bm{P}_F) \bm{x},
\end{equation}
where
\begin{equation}
	\bm{P}_F = \bm{F} \left(\bm{F}^\dagger \bm{F}\right)^{-1}\bm{F}^\dagger
\end{equation}
is the projection matrix onto the subspace of $\bm{F} = {[{\bm{p}}_{f_1}, \dots, {\bm{p}}_{f_S}]} \in \mathbb{C}^{N \times S}$.

For the problem at hand, a computationally and space efficient implementation is based on a suitable orthogonalization procedure. In particular, let $\Omega = \Omega_1 \cup \Omega_2 \cup \dots \cup \Omega_K$ be the union of the stop frequency bands, with $\bm{Q}_N = [\bm{Q}^1_N, \dots, \bm{Q}^K_N]$ {the} set of orthonormal steering vectors {whose normalized frequencies lie in} $\Omega$, starting from $\bm{c}_0$, the method iteratively calculates a vector orthogonal to $\Omega$ as
\begin{equation}
	\bm{c} = \bm{c}_0 - { \sum_{{f_i \in \Omega}} \bm{p}_{f_{i}} \bm{p}_{f_{i}}^\dagger} \bm{c}_0.
\end{equation}

\begin{algorithm}[t]
	\caption{Projection-Based Spectral Notching design}
	\label{alg:proj_notch}
	\KwIn{$N$, $\{\Omega_k\}_{k=1}^K$, {$\bm{c}_0$}}
	set $\bm{c} = \bm{c}_0$\;
	{set $\Omega = \Omega_1 \cup \Omega_2 \cup \dots \cup \Omega_K$}\;
	{\ForEach{$f \in \Omega$}{
	define	$\bm{p}_{f_{i}} = \frac{1}{\sqrt{N}} [1, \exp(-j2\pi f_{i}), \dots, \exp(-j2\pi(N-1) f_{i})]^\mathrm{T}$\;	
	  compute $\bm{c} = \bm{c} - \bm{p}_{f_{i}} \bm{p}_{f_{i}}^\dagger \bm{c}_0$\;
	 }}
	\KwOut{$\bm{c}$}
\end{algorithm}
While this method {(streamlined in \textbf{Algorithm~2})} has the advantage of being computationally efficient, it provides less control over the waveform characteristics than the QCQP-based approach. However, it {is} also worth pointing out that this procedure provides a solution to problem $\mathcal{P}$ when {$L=1$ and} $E_I^k = 0, \; k=1, \dots K$, i.e., when the acceptable levels of disturbance are set to 0.
Also, {note} that this approach preserves the Gaussian nature of the signal. In other words, if the reference signal $\bm{c}_0$ is drawn from a Gaussian distribution, then the resulting waveform will also follow the same distribution.

Before concluding, two important remarks are necessary.
\begin{enumerate}
	\item[a)] These techniques can be used to design both continuous jamming interference and to induce noise cover pulses preserving friendly RF systems.
	\item[b)] These methodologies can be easily modified to produce a multiple spot jamming signal capable of covering only some tactical regions for the considered bandwidth.
\end{enumerate}

\section{QUANTIZATION AND ITS EFFECT ON SPECTRAL NOTCHES}
{This section focuses on the impact of the quantization involved in the analog conversion of the jamming waveforms}, with a particular focus on the spectral alterations resulting from reduced DAC resolution.

Quantization is the process of mapping a large set of input values to a smaller set {of discrete output values}. This process arises in modern radio and radar devices, where the digital signals, synthesized numerically via a specific design process {that leverages} a large number of bits, must be {converted into continuous analog signals} using a DAC with a (generally) lower bit resolution {before being transmitted through the analog front-end.} This conversion involves {(as first step)} representing the digital input, which is typically in a high-resolution binary format, with a finite number of discrete levels determined by the DAC resolution. The higher the bit resolution, the more precise the representation of the original digital signal~\cite{toulson2012fast}.

{Notably}, quantization could affect the fidelity of the transmitted waveform, whereby small changes in the signal amplitude may be lost in the quantization noise~\cite{smith1997scientist}. {In particular, for spectrally-shaped waveforms, a reduction in quantization resolution could potentially result in a modification of the waveform spectral behavior, which might lead to a failure of the intended objective of avoiding {interference} with friendly emitters.}

At the transmission stage,  both the real and imaginary components of the complex baseband signal {(whose values are assumed within the range $[-1,1]$ {and described with an infinite precision})} {undergo} an implicit quantization operation, {in order to represent the digital sequence using the same} number of bits employed by the DAC, with a resulting truncation (or rounding) of its least significant bits. Precisely, {to study the quantization impact on the spectral characteristics of the devised signals, the} real and the imaginary component of each element  $c(i), i = 1, \dots, N, $ of the synthesized waveform $\bm{c}$ is quantized by representing it using a finite set of $2^b$ values, where $ b $ denotes the number of DAC bits. This operation can be formally expressed as
\begin{equation}\label{eq:quant}
	{\check{c}_b(i) = Q_b({\rm{Re}}\{c(i)\}) + j Q_b({\rm{Im}}\{c(i)\})},
\end{equation}
where $\check{c}_b(i)$ {represents the term} $c(i)$ {whose real and imaginary parts are quantized with $b$ bits}. In this context, the quantization error between the $i$th element ($i=1, \dots, N$) of the original waveform $ \bm{c} $ and the quantized one $\bm{\check{c}_b}$, is defined as
\begin{equation}\label{eq:quantization_error}
	\epsilon(i) = c(i) - \check{c}_b(i).
\end{equation}
Notably, {neglecting the overload distortion,} the real and imaginary parts of $\epsilon(i)$ follow {approximately} a uniform distribution~\cite{smith1997scientist}.
{Furthermore, since the range of the real/imaginary part} of the quantization error values decreases as the number of bits increases, the variance of the {(real/imaginary)} quantization error $ \sigma^2_\epsilon $ is inversely proportional to the square of the {number of} quantization levels (i.e., a function of $ b $), {namely}
\begin{equation}\label{eq:variance_th}
	\sigma^2_\epsilon \approx \frac{\Delta_b^2}{12} = \frac{1}{12} \left(\frac{{2}}{2^b}\right)^2,
\end{equation}
where $\Delta_b = {2/(2^b)}$ is the quantization step size when $b$ bits are employed {to discretize the interval $[-1,1]$}.

In the context of cognitive waveforms designed for spectral compatibility, a figure of merit is the decrease in the {depth of} spectral notches {resulting} by the quantization process. This is an important factor to assess the impact of quantization, as energy spilling into the frequency stop-bands could potentially {induce} interference {on} friendly emitters.

\section{NUMERICAL SIMULATION}\label{section:numerical_sim}
In this section, the waveforms synthesized with \textbf{Algorithm 1} and \textbf{Algorithm 2} are analyzed in terms of spectral {features} and autocorrelation behaviour. At the synthesis stage, a jamming bandwidth of $20$ MHz and a {sequence} length $N = 10^5$, corresponding to the design of a waveform with a duration of 5 $m$s, are {considered}; furthermore, 3 stop-bands located in the (baseband) intervals $[-4,-2]$ MHz, $[4,5]$ MHz, and $[8,9]$ MHz have been {introduced} for spectral shaping. A constant modulus sequence with random-phases is used as reference signal $\bm{c}_0$, whose $i$th element is given by 
\begin{equation}
	c_0(i) = (1/\sqrt{N}) e^{j 2 \pi \phi_i}, i=1,\dots, N,
\end{equation}
where $\phi_i$ are independent and identically distributed random variables with a uniform probability density function over $[0,1]$.

By employing \textbf{Algorithm 1} with {average} notch depths set to 60 dB, {the two waveforms synthesized assuming $\bar{N} = 1000$ or $\bar{N} = 5000$ and $W=\bar{N}/2$ are referred to in the following as WF-QCQP(5000) and WF-QCQP(1000). The waveform designed via \textbf{Algorithm 2} is indicated as WF-PROJ.}

\begin{figure}[t]
	\centering
	\includegraphics[trim={30 0 40 30}, clip, width=0.98\linewidth]{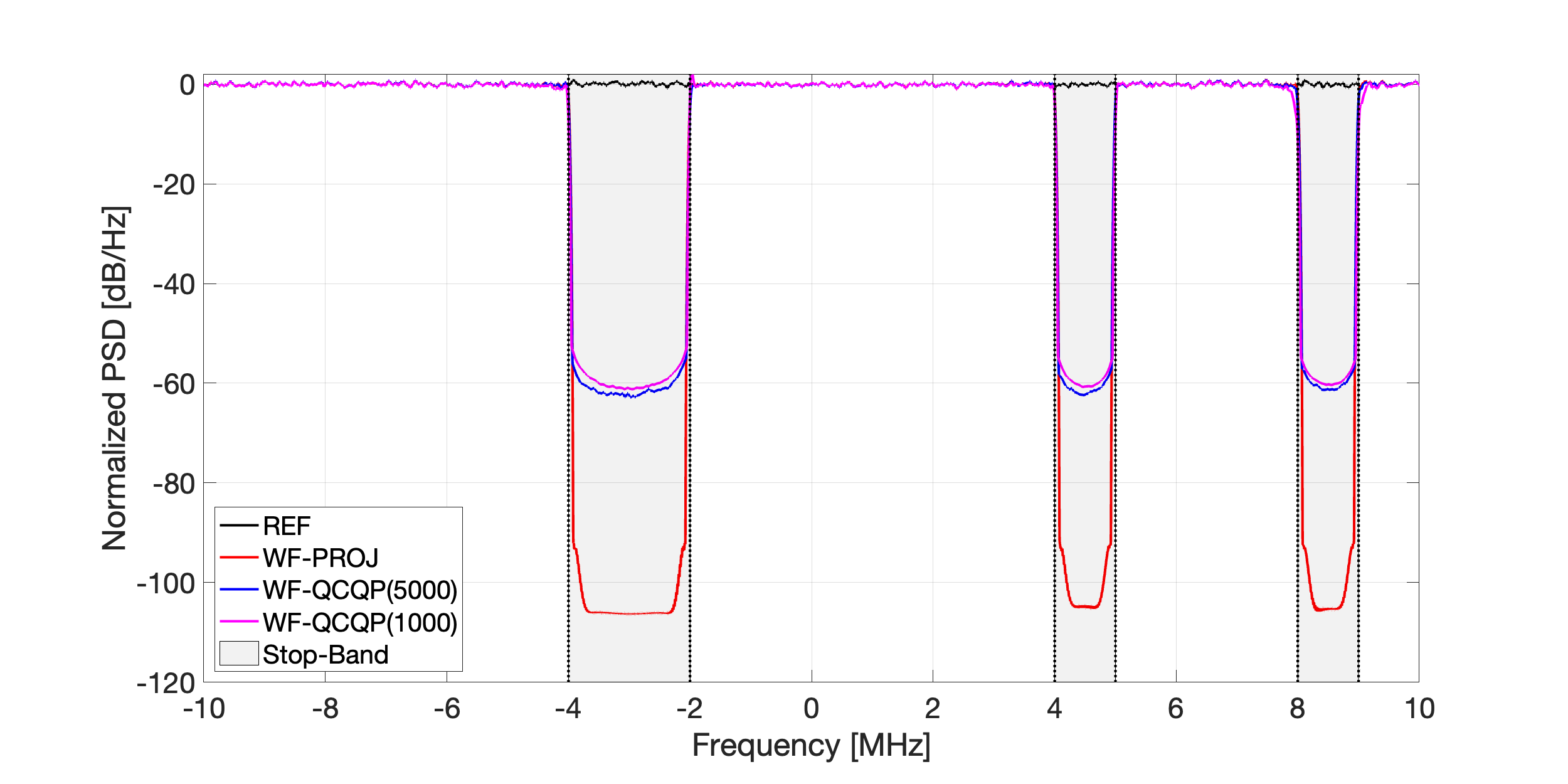}
	\caption{{PSD estimate} of the reference waveform and the synthesized ones via \textbf{Algorithm 1} (using block-size either of 1000 or 5000) and \textbf{Algorithm 2}. The stop-bands are depicted in light gray. {The PSD are estimated using the Welch method (considering segments of 1000 samples with 50$\%$ overlap and weighted with an Blackman-Harris window) and normalized to the mean value of the reference waveform PSD}.}
	\label{fig:periodogram_sim} 
\end{figure}

\begin{figure}[t]
	\centering
	\includegraphics[trim={100 35 100 30}, clip, width=0.95\linewidth]{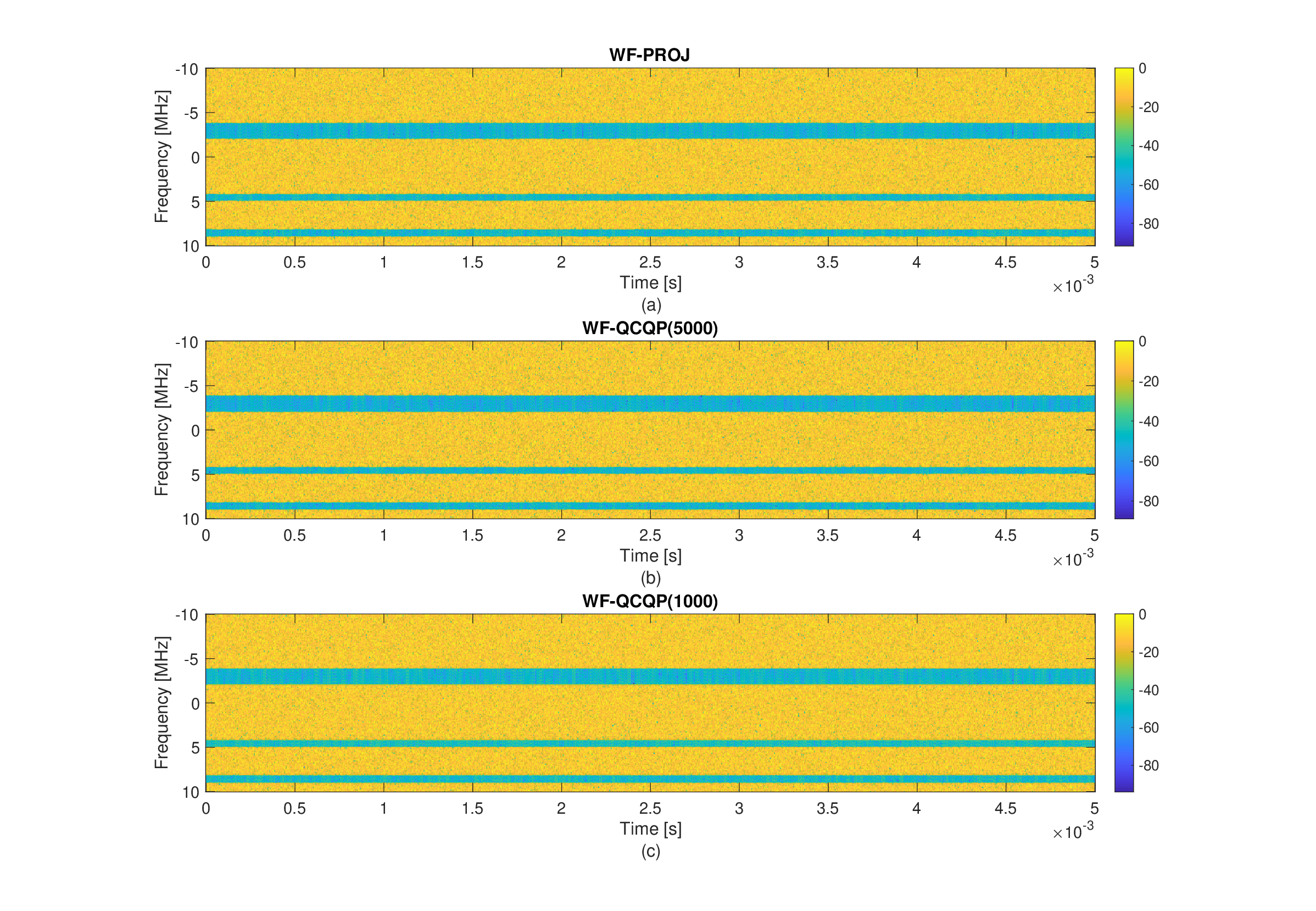}
	\caption{Spectrogram (normalized to the maximum value) of the waveforms synthesized using \textbf{Algorithm 1} (using $\bar{N}=1000$ and $\bar{N}=5000$) and \textbf{Algorithm 2}, computed on signal segments of 10 $\mu$s.}
	\label{fig:spectrogram_sim} 
\end{figure}

\begin{figure}[t]
	\centering
	\includegraphics[width=0.95\linewidth]{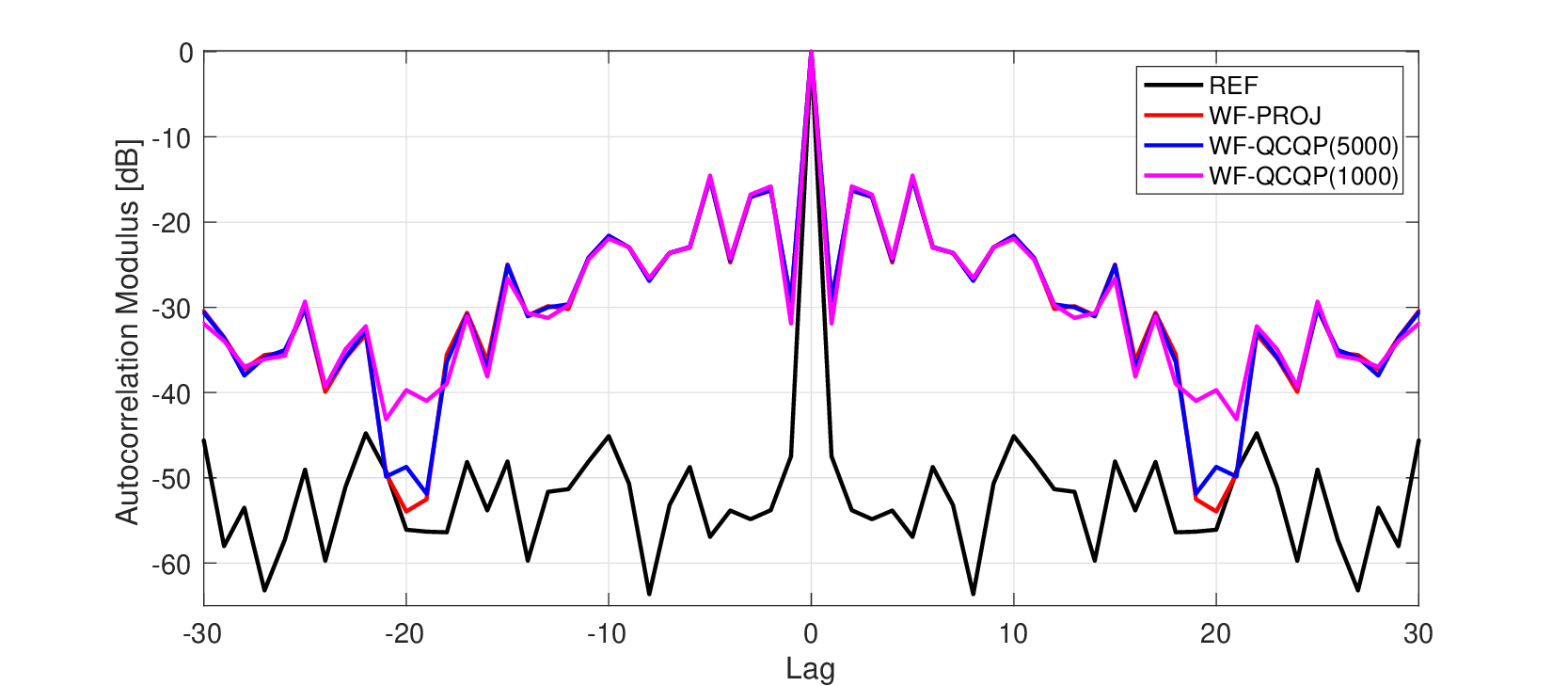}
	\caption{{Normalized} autocorrelation of the synthesized waveforms computed on the whole sequence. The plot is zoomed-in for ease of visualization.}
	\label{fig:autocorr_sim} 
\end{figure}
Fig.~\ref{fig:periodogram_sim} reports the power spectral density (PSD) of the reference waveform $\bm{c}_0$ (hereafter referred to as REF) alongside {those corresponding to the} aforementioned {counterparts}, estimated using the Welch method (considering segments of 1000 samples with 50$\%$ overlap and weighted with an Blackman-Harris window) and normalized to the mean PSD value of the reference waveform. The {figure} clearly shows that all three designed waveforms exhibit spectral notches perfectly lined-up within the desired stop-bands. On the other side, as expected the reference waveform looks like {a} frequency flat noise over {all} the operational bandwidth.
As to the waveforms generated using \textbf{Algorithm 1}, it is evident that regardless {of} the used block length (i.e., $1000$ or $5000$), the {notches} depth is approximately $61$ dB, with slightly higher values near the boundaries of the stop-bands but always {deeper than} $50$ dB. Notably, the spectral notch level of the waveform obtained with \textbf{Algorithm 2} is deeper than $60$ dB, with PSD values {in the order of $-110$~dB within the stop bands and -90 dB near the boundaries}.

Fig.~\ref{fig:spectrogram_sim} shows the spectrogram (normalized to the maximum value) of the three designed {signals}, computed over segments {with} 200 samples (equivalent to {a} 10 $\mu$s temporal window given the sampling frequency) with a $50\%$ overlap between segments.
An inspection of the plots highlights the presence of {stationary} spectral notches, which are visible even over short {segments with} 10 $\mu$s duration. It is also noteworthy that the notches are present within the designated stop-bands, while in other regions, the spectrum appears quite {flat} with values close to 0 dB.

To quantify the noise-like behaviour of the {designed spectrally-notched jamming signals}, Fig.~\ref{fig:autocorr_sim} {compares} {the modulus of their} autocorrelation {functions (normalized to their maximum) with that of the reference signal}, limited in a neighborhood of the 0-lag for ease of visualization. It can be seen from the curves that regardless of the employed procedure, the obtained {jamming sequence} is characterized by an autocorrelation peak sidelobe level (PSLL) of $-14.5$ dB, while for the reference waveform it is equal to $-44.8$ dB, with an actual offset of about $30$ dB.

\begin{figure}[t]
	\centering
	\includegraphics[width=0.95\linewidth]{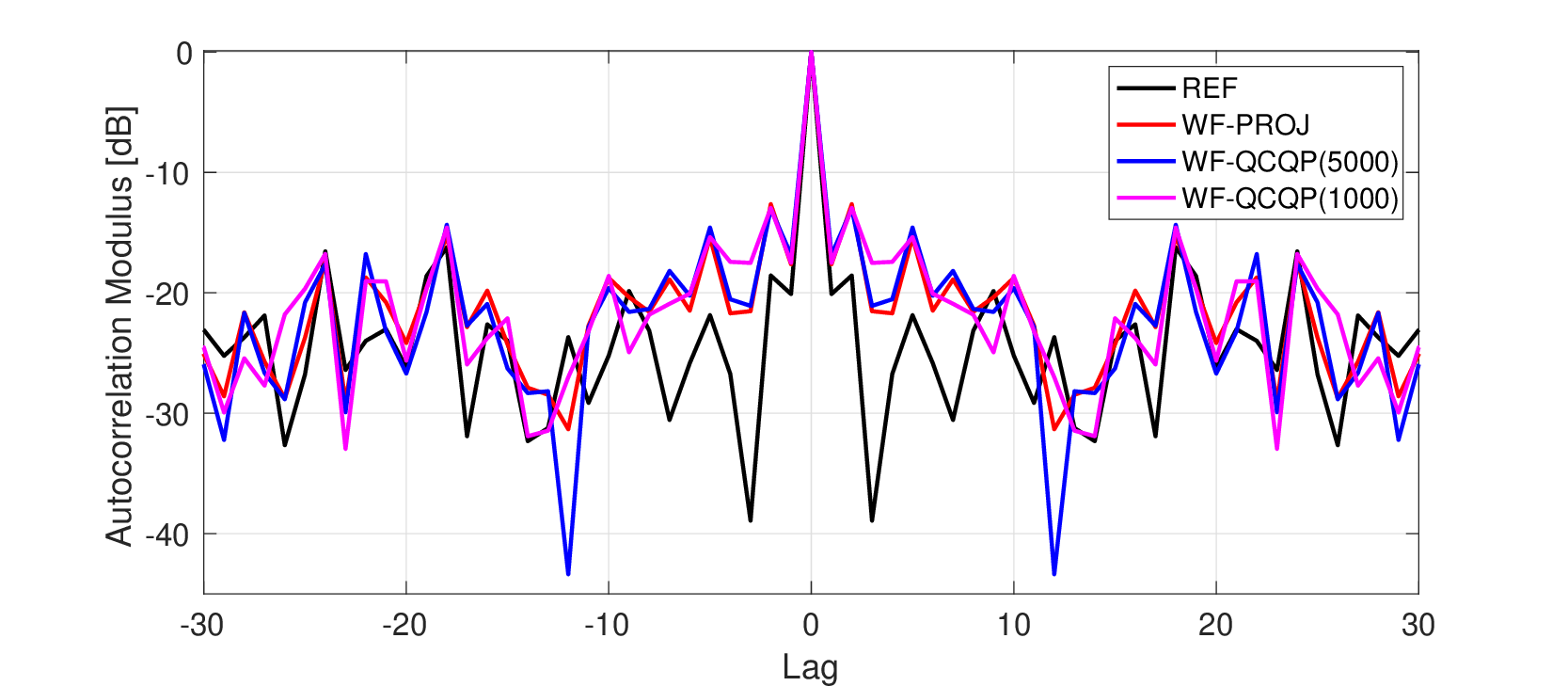}
	\caption{{Normalized} autocorrelation of the synthesized waveforms computed on a segment of 200 samples, corresponding to a 10 $\mu$s long {sequence}. The plot is zoomed-in for ease of visualization.}
	\label{fig:autocorr_chunk_sim} 
\end{figure}

In contrast, Fig.~\ref{fig:autocorr_chunk_sim} shows the {normalized} autocorrelation {computed considering a small segment of 200 signal} samples, corresponding to 10 $\mu$s. For this case, {all the synthesized waveforms achieve a PSLL of about $-12.9$ dB with a gap with respect to the reference in the order of} $7$ dB.

In both {the} analyses, there are no significant differences between the waveforms designed {via} \textbf{Algorithm 1} (regardless of block size) and \textbf{Algorithm 2}. This is because the imposed notch level ($60$ dB) is so tight that {the  number of degrees of freedom that can be utilized by the design procedure described in \textbf{Algorithm 1} do not make possible to further} minimize its objective function, i.e., maximize the similarity {with} the reference waveform, while satisfying the spectral constraints. Therefore, in this scenario, the two approaches yield very similar {sequences}.
As previously highlighted, {in the case of a single block, i.e., $L=1$,} if lower notch depths were imposed, e.g., {larger than} $200$ dB, the two approaches would generate {exactly} the same {signal}.

\begin{figure}[t]
	\centering
	\includegraphics[trim={30 0 40 10}, clip, width=0.95\linewidth]{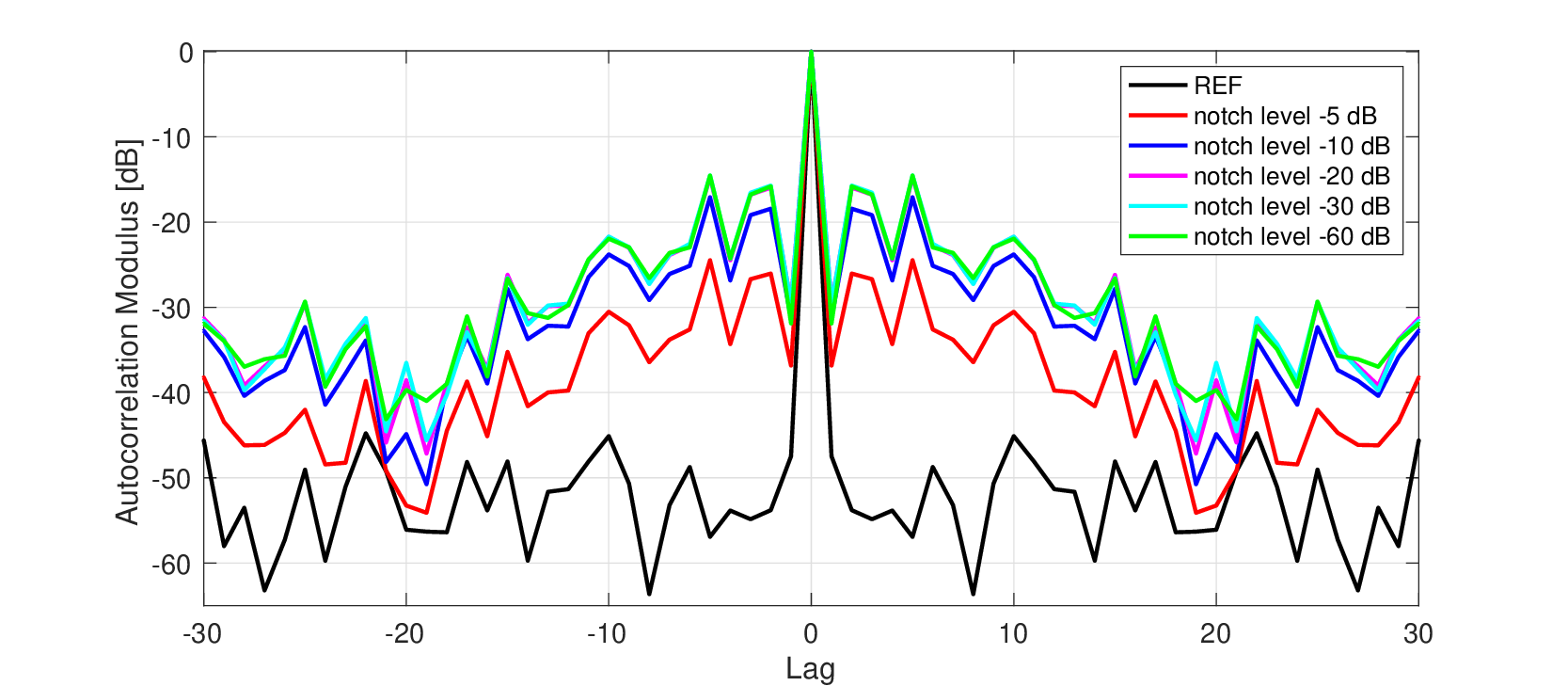}
	\caption{{Normalized}  autocorrelation of the waveforms synthesized using \textbf{Algorithm 1} with block-size 1000 and considering notch {depths} of $5$ dB, $10$ dB, $20$ dB, $30$ dB, $60$ dB.}
	\label{fig:autocorr_qcqp_1000} 
\end{figure}

To further illustrate the trade-off between notch depth and similarity with the reference sequence, Fig.~\ref{fig:autocorr_qcqp_1000} presents the {normalized} autocorrelations of the waveforms synthesized via \textbf{Algorithm 1} with a block length of 1000 and notch depth {ranging} from $5$ dB to $60$ dB. The results demonstrate that as the notch depth increases, the resulting PSLL also increases. Additionally, for the considered scenario, there is no degradation in autocorrelation {features} observed for notch {depths below} $20$ dB.

\subsection{Quantization Effects}
In the following, the effect of quantization on spectrally-shaped jamming waveforms is numerically evaluated considering, as a case study, the frequency-selective {non-coherent} waveform synthesized using \textbf{Algorithm 2}.

\begin{figure}[t]
	\centering
	\includegraphics[width=0.99\linewidth]{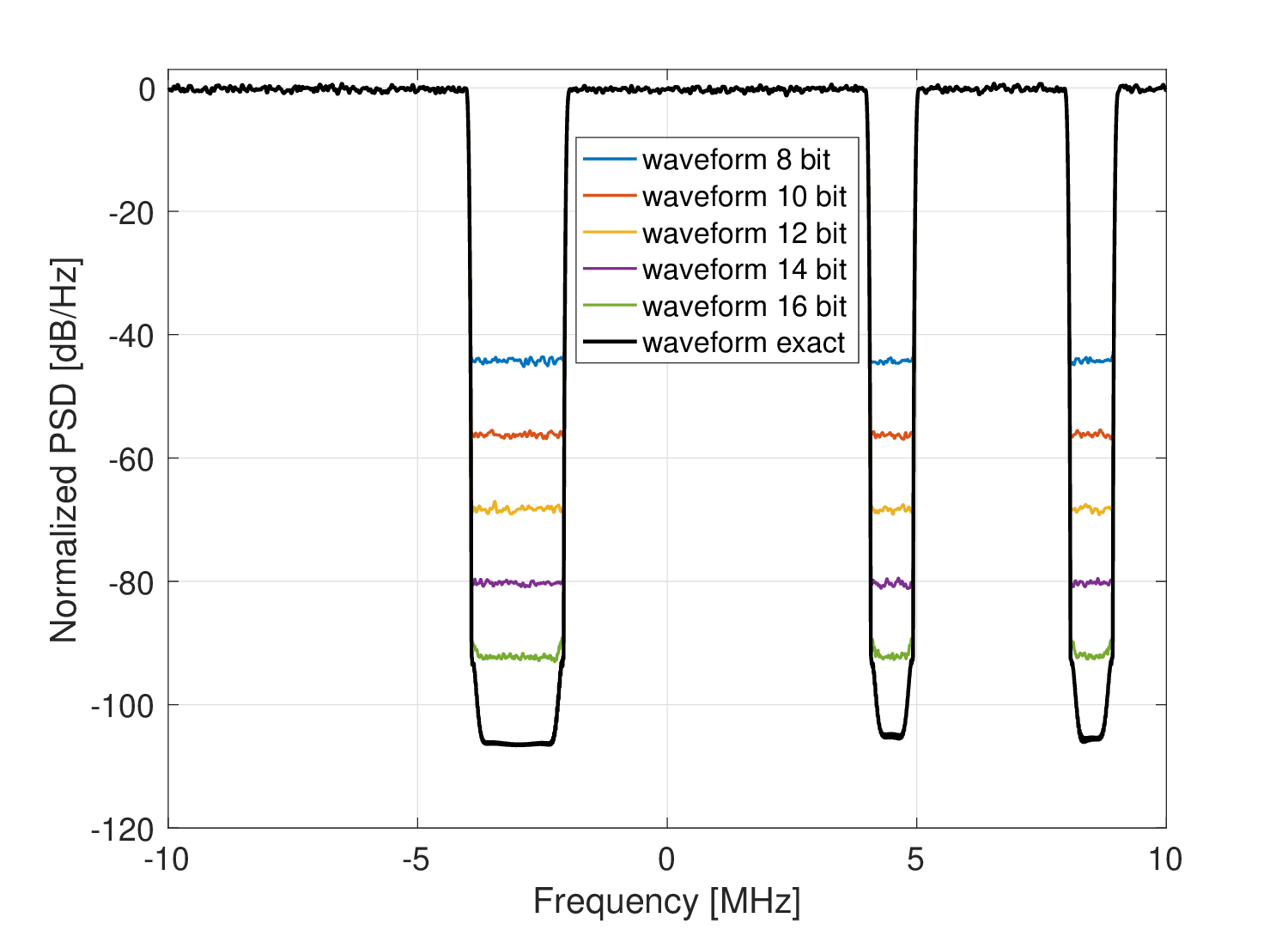}
	\caption{PSD of the selective {non-coherent} jamming waveform synthesized using \textbf{Algorithm 2} and its versions quantized with 8, 10, 12, 14, and 16 bits.}
	\label{fig:barrage_proiezione}
\end{figure}

{Assuming the same scenario as in Fig.~\ref{fig:periodogram_sim}, Fig.~\ref{fig:barrage_proiezione} compares the normalized PSDs estimated using the waveform designed via \textbf{Algorithm 2} with those estimated using its quantized versions {where} 8, 10, 12, 14, {or} 16 bits {are used to represent the real and the imaginary parts of the samples, respectively}. Specifically, the PSDs are estimated using the Welch method employing segments of 1000 samples (with 50\% overlap) and weighted with a Blackman-Harris window.}

Inspection of the curves confirms that the notch level of the waveform increases as the number of bits decreases. In particular, the synthesized waveform is characterized by a notch level of -110 dB, while there is a noticeable increase in {mentioned metric} of the quantized waveforms that is inversely proportional to the employed resolution. The 16-bit quantization {allows for a better approximation of the ideal waveform spectral shape}, still with a notch depth of 92 dB. However, with 8-bit quantization, the spectral notch is significantly shallower, with a resulting level of -44 dB, which is inadequate to fulfill the spectral coexistence {specifications}.

The observed quantization effect is also confirmed by the energy of the difference between the synthesized waveform and its quantized version, i.e., $\|\bm{c} - \bm{\check{c}_b}\|^2$, whose values are reported in Table~\ref{tab:norm_difference}. Not surprisingly,  the results pinpoint that as the number of bits increases, the energy of the difference decreases, i.e., the quantized waveform better approximates the synthesized one.

\begin{figure*}[t] 
	\centering
	\subfloat[]{%
		\includegraphics[width=0.48\linewidth]{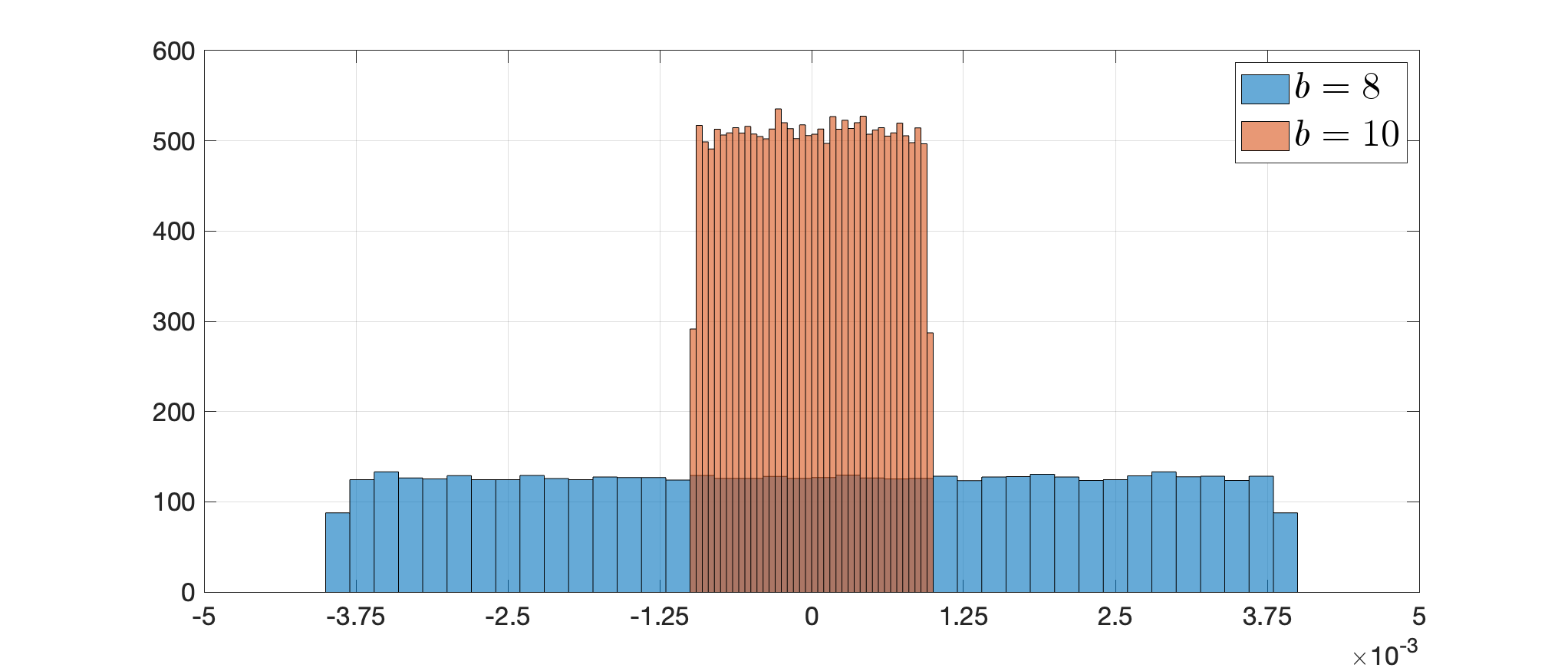}} \hfil
	\subfloat[]{%
		\includegraphics[width=0.48\linewidth]{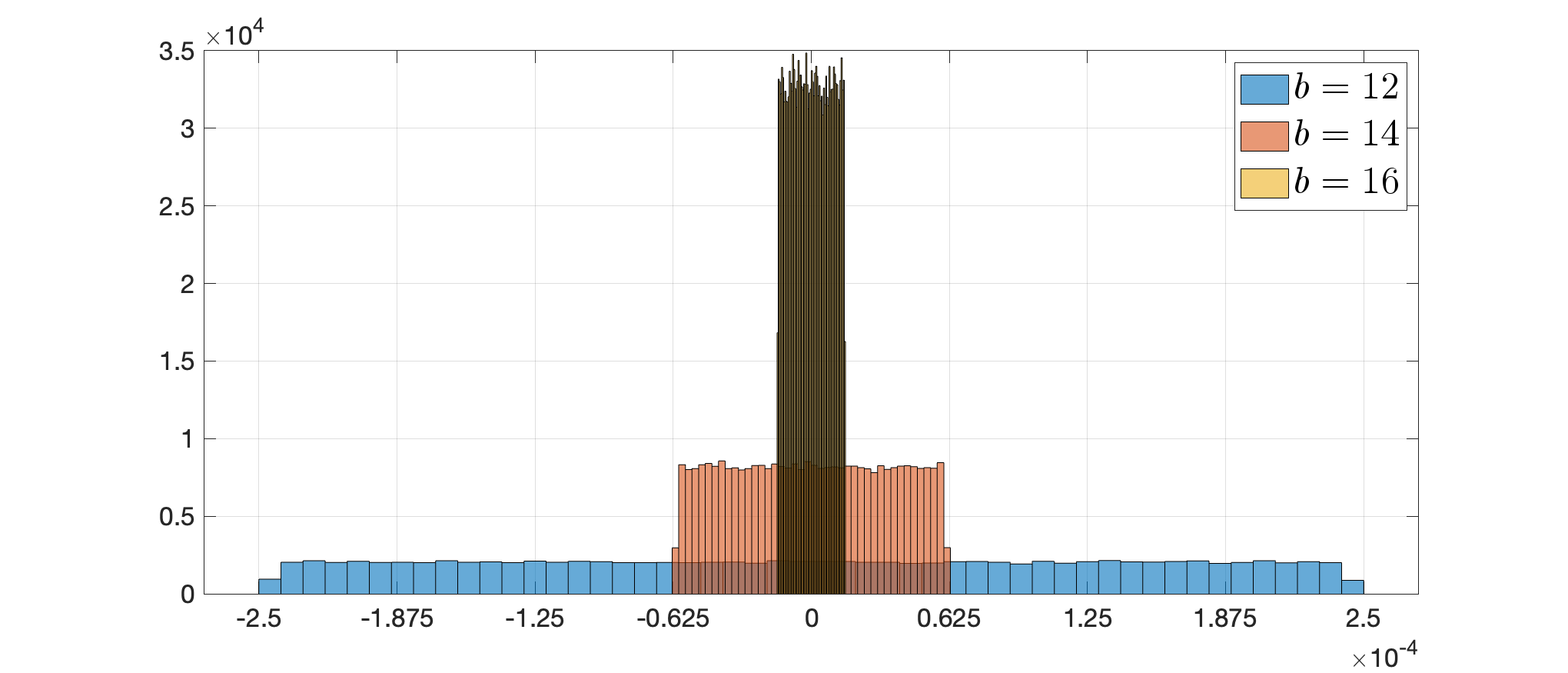}} \hfil
	\caption{\blue{Empirical distributions of the real part of the quantization error for different values of $b$: (a) $b=\{8,10\}$, (b) $b=\{12,14,16\}$.}}
	\label{fig:hist_quant_error} 
\end{figure*}

{Finally}, in Table~\ref{tab:variance_comparison}, the theoretical values of the quantization error variance~\eqref{eq:variance_th} are compared with the estimated ones (considering the real part of the quantization error), revealing that there is a good agreement between the numerical and the theoretical values, corroborating the goodness-of-fit of the considered {additive} model {for the quantization error}. The corresponding empirical distribution of the real part of the quantization error~\eqref{eq:quantization_error} {is} reported in Fig.~\ref{fig:hist_quant_error}, considering 8, 10, 12, 14, {and} 16 bits for the digital quantization. The plots {show} that, regardless of the number of bits, the real part of the quantization error is uniformly distributed, with parameters in line with the variance values reported in Table~\ref{tab:variance_comparison}. Notably, the same estimated variance and empirical distributions are {also} obtained by considering the real part of the quantization error.

In summary, a quantization process with fewer number of bits result in spectral characteristics {degradation of} the signal, with {deteriorated} spectral notches {as} compared {with} the design levels. This potential alteration can lead to unintended interference within the frequencies occupied by friendly systems.
{The conducted study sheds light on} the critical impact of DAC bit resolution on the performance of spectrally-shaped waveforms.

\begin{table}[t]
	\centering
	\caption{Energy difference between exact and quantized waveform.}
	\begin{tabular}{cc}
		\hline
		\hline
		\textbf{\# of Bits} &  $\|\bm{c} - \bm{\check{c}_b}\|^2$\\
		\hline
		\textbf{8} & \num{1.0345} \\
		\textbf{10} &  \num{0.063391}   \\
		\textbf{12} &  \num{0.0039831}  \\
		\textbf{14} &  \num{0.00024859}  \\
		\textbf{16} &  \num{1.5523e-05} \\
		\hline
		\hline
	\end{tabular}
	\label{tab:norm_difference}
\end{table}

\begin{table}[t]
	\centering
	\caption{Estimated vs theoretical variance of the quantization error.}
	\begin{tabular}{ccc}
		\hline
		\hline
		\textbf{\# of Bits} & \textbf{Estimated Variance} & \textbf{Th. Variance~\eqref{eq:variance_th}} \\
		\hline
		\textbf{8}  & \num{5.167e-6} & \num{5.0863e-06} \\
		\textbf{10} & \num{3.191e-7} & \num{3.1789e-07} \\
		\textbf{12} & \num{1.989e-8} & \num{1.9868e-08} \\
		\textbf{14} & \num{1.242e-9} & \num{1.2418e-09} \\
		\textbf{16} & \;\;\num{7.761e-11} & \;\;\num{7.7610e-11} \\
		\hline
		\hline
	\end{tabular}
	\label{tab:variance_comparison}
\end{table}

\color{black}
\subsection{Effect of Increased $K$}\label{subsection:K}
\begin{figure}[t] 
	\centering
		\includegraphics[width=0.99\linewidth]{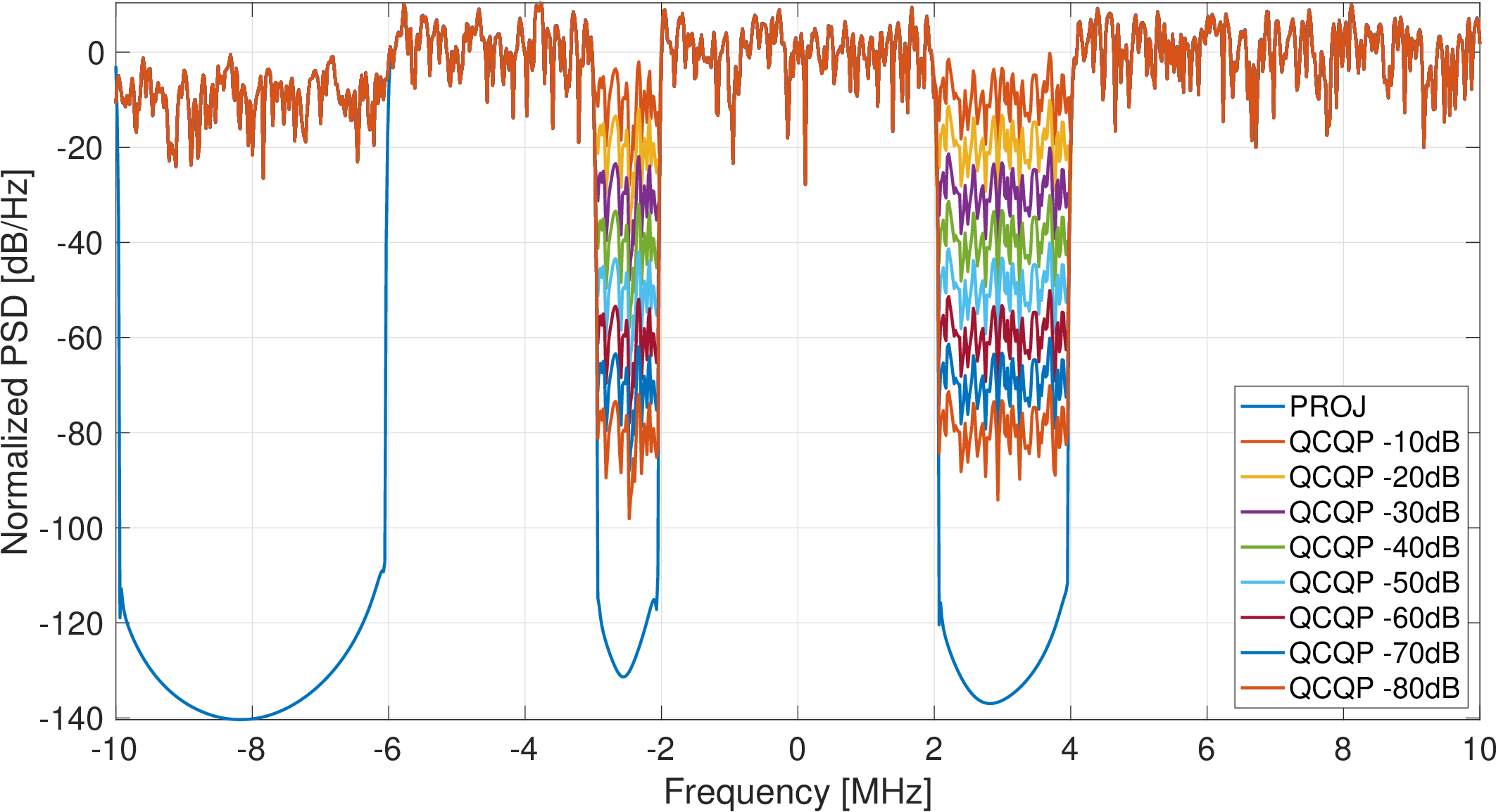}
	\caption{\blue{PSD estimates of the synthesized cognitive waveforms for Case 1 of Subsection~\ref{section:numerical_sim}-\ref{subsection:K}. The PSDs are estimated using the Welch method (applied to each waveform using a Blackman-Harris tapering) and normalized to their mean value.}}
	\label{fig:fig_K_3_N1000} 
\end{figure}

\begin{figure}[t] 
	\centering
		\includegraphics[width=0.99\linewidth]{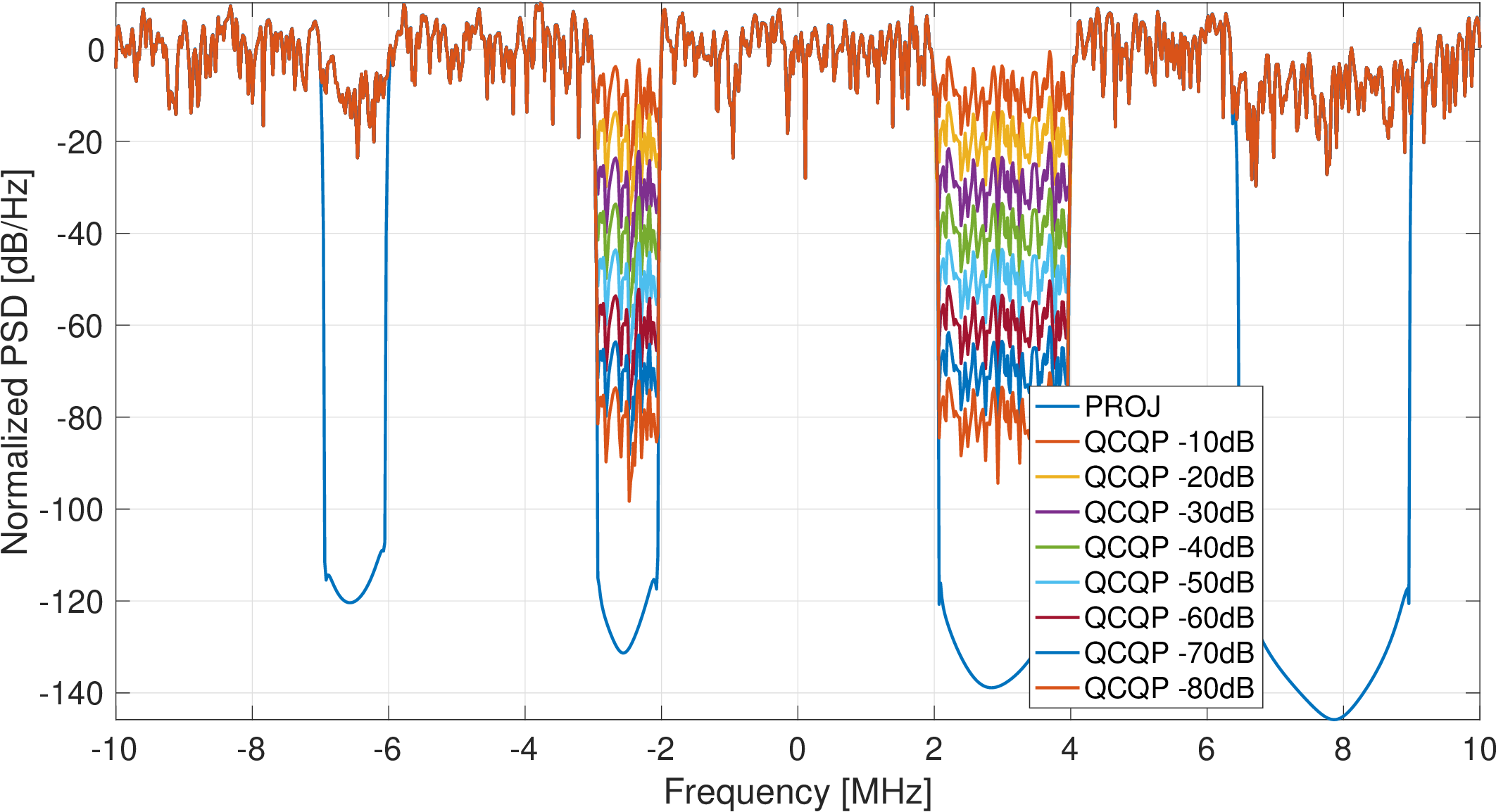} \hfil \\
	\caption{\blue{PSD estimates of the synthesized cognitive waveforms for Case 2 of Subsection~\ref{section:numerical_sim}-\ref{subsection:K}. The PSDs are estimated using the Welch method (applied to each waveform using a Blackman-Harris tapering) and normalized to their mean value.}}
	\label{fig:fig_K_4_N1000} 
\end{figure}

\begin{figure}[t] 
	\centering
		\includegraphics[width=0.99\linewidth]{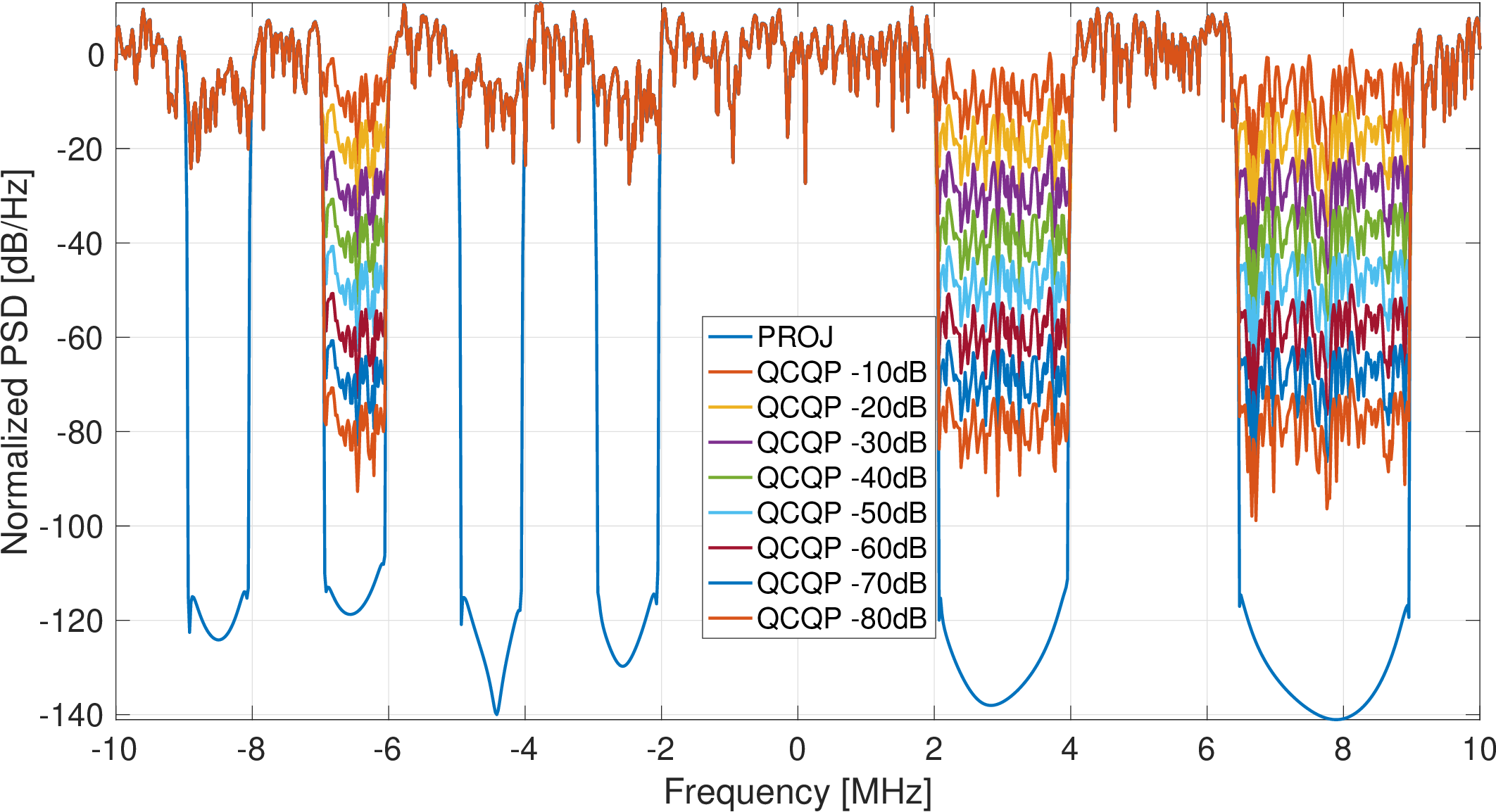} \hfil \\
	\caption{\blue{PSD estimates of the synthesized cognitive waveforms for Case 3 of Subsection~\ref{section:numerical_sim}-\ref{subsection:K}. The PSDs are estimated using the Welch method (applied to each waveform using a Blackman-Harris tapering) and normalized to their mean value.}}
	\label{fig:fig_K_6_N1000} 
\end{figure}

The effect of the parameter $K$ is investigated considering three simulation scenarios involving 3, 4, and 6 friendly emitters operating in the environment, having different spectral supports and tolerable interference level (consequently notch depth) as follows.
\begin{itemize}
	\item Case 1: $K = 3$ friendly emitters operating over the frequency intervals $[-10, -6]$ MHz, $[-3, -2]$ MHz, and $[2, 4]$ MHz, with a notch depth of 10 dB for the first emitter and notch depths ranging from 10 dB to 80 dB for the others.
	
	\item Case 2: $K = 4$ friendly emitters operating over the frequency intervals  $[-7,-6]$ MHz, $[-3,-2]$ MHz, $[2,4]$ MHz, and $[6.5,9]$ MHz; first and last notches having depth $10$ dB and the others ranging from 10 dB to 80 dB.
	\item Case 3: $K = 6$ friendly emitters operating over the frequency intervals $[-9,-8]$ MHz, $[-7,-6]$ MHz MHz,$[-5,-4]$ MHz, $[-3,-2]$ MHz, $[2,4]$ MHz, and $[6.5, 9]$ MHz; first, third, and fourth notches having depth $10$ dB and the others ranging from 10 dB to 80 dB.
\end{itemize}
In this regard, the cognitive waveforms (considering a bandwidth of 20 MHz) are composed of a single block ($L = 1$) with $N = 1000$ samples.
The normalized PSDs of the cognitive waveforms designed in the aforementioned case studies are illustrated in Figs.~\ref{fig:fig_K_3_N1000}, \ref{fig:fig_K_4_N1000}, and \ref{fig:fig_K_6_N1000}, respectively.

Inspection of the figures confirms that the QCQP-based waveform design (\textbf{Algorithm 1}) successfully achieves the desired spectral shaping, with spectral notches meeting the specified depth requirements. In contrast, \textbf{ Algorithm 2} allows controlling only over the notch locations, not their depths, thus limiting design flexibility. Nonetheless, in frequency bands outside the notches (i.e., the pass-bands), both methods produce waveforms with {almost} identical PSD.

\begin{figure}[t] 
	\centering
		\includegraphics[width=0.99\linewidth]{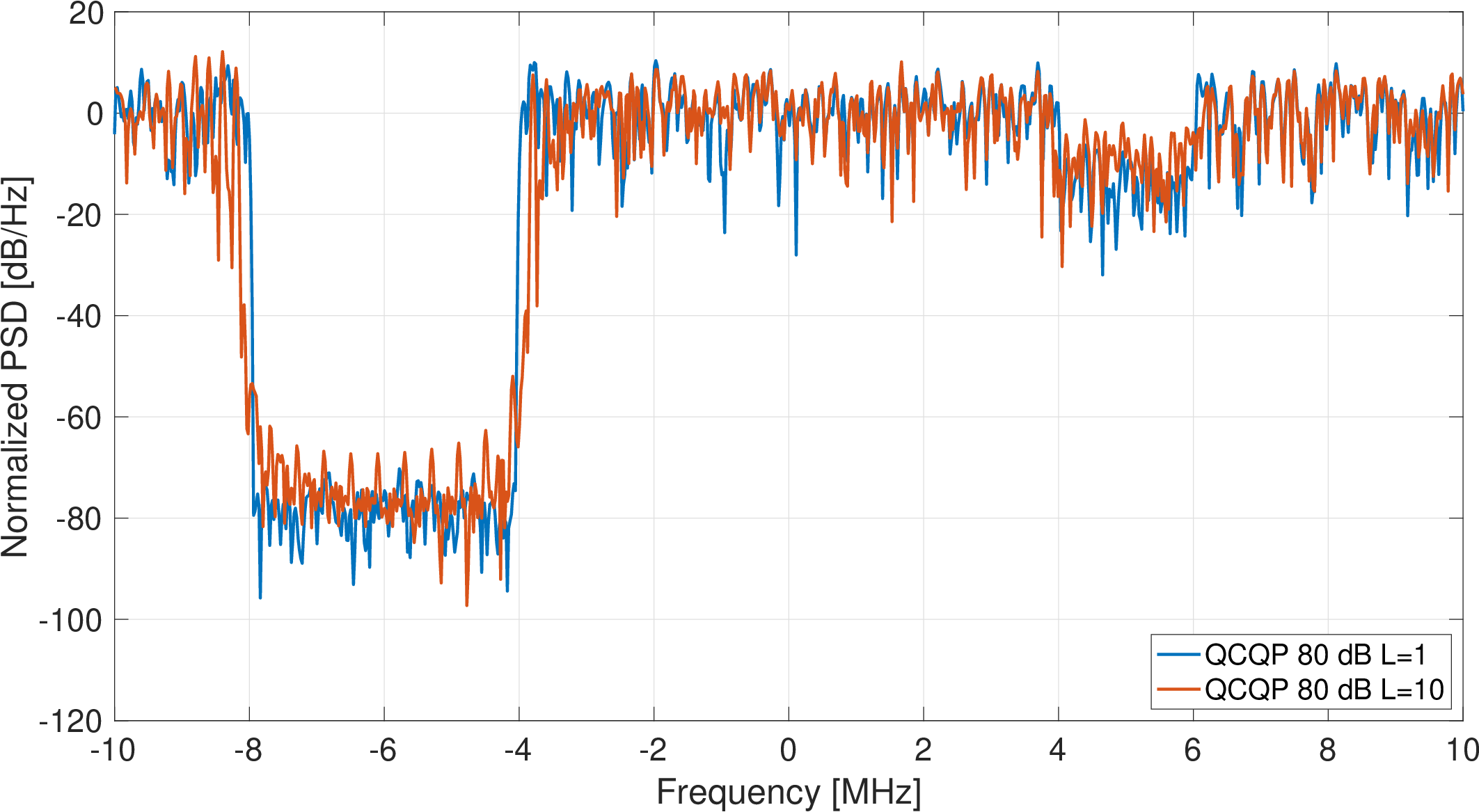}
	\caption{\blue{PSD estimates of QCQP-based synthesized cognitive waveforms for $K=2$, $N=1000$, and $L \in \{1,10\}$. The PSDs are estimated using the Welch method (applied to each waveform using a Blackman-Harris tapering) and normalized to their mean value.}}
	\label{fig:diffL} 
\end{figure}

\subsection{Effect of Increased $L$}\label{sub:incr_L}
A scenario with $K=2$ friendly emitters occupying the frequency intervals $[-8, -4]$ MHz and $[4, 6]$ MHz, is considered to assess the impact of the block length. A waveform with $N=1000$ samples and 20 MHz bandwidth is synthesized using \textbf{Algorithm 1} under two configurations: the former using a single block, i.e., $L=1$, while the latter employing $L=10$ blocks, each composed of 100 samples. For both configurations, the notch depth in the first stop band is 80 dB, while in the other is 10 dB.

The resulting PSD estimates, reported in Fig.~\ref{fig:diffL}, reveal that while both designs exhibit similar behavior in the pass-bands, employing a larger number of blocks results in a degradation of the spectral notches characteristics. Additionally, the transition between the pass-band and stop-band is smoother in the multi-block case, indicating a degradation in spectral selectivity. These results underscore the importance of carefully balancing design flexibility and spectral accuracy.

\color{black}

\section{EXPERIMENTAL ANALYSIS}

\begin{figure}[ht]
	\centering
	\includegraphics[width=0.99\linewidth]{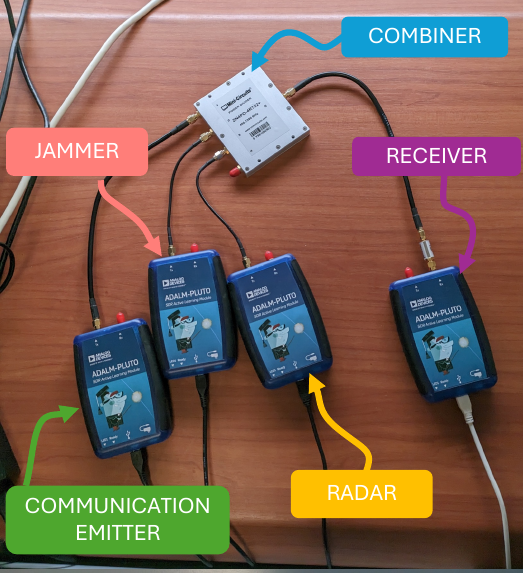}
	\caption{{Illustration} of the experimental setup {comprising four PLUTO-SDR and a Mini-Circuits combiner}.}
	\label{fig:test-bed} 
\end{figure}

\begin{figure*}[t]
	\centering
	
	\subfloat[]{%
		\includegraphics[trim={35 0 45 10}, clip,width=0.48\linewidth]{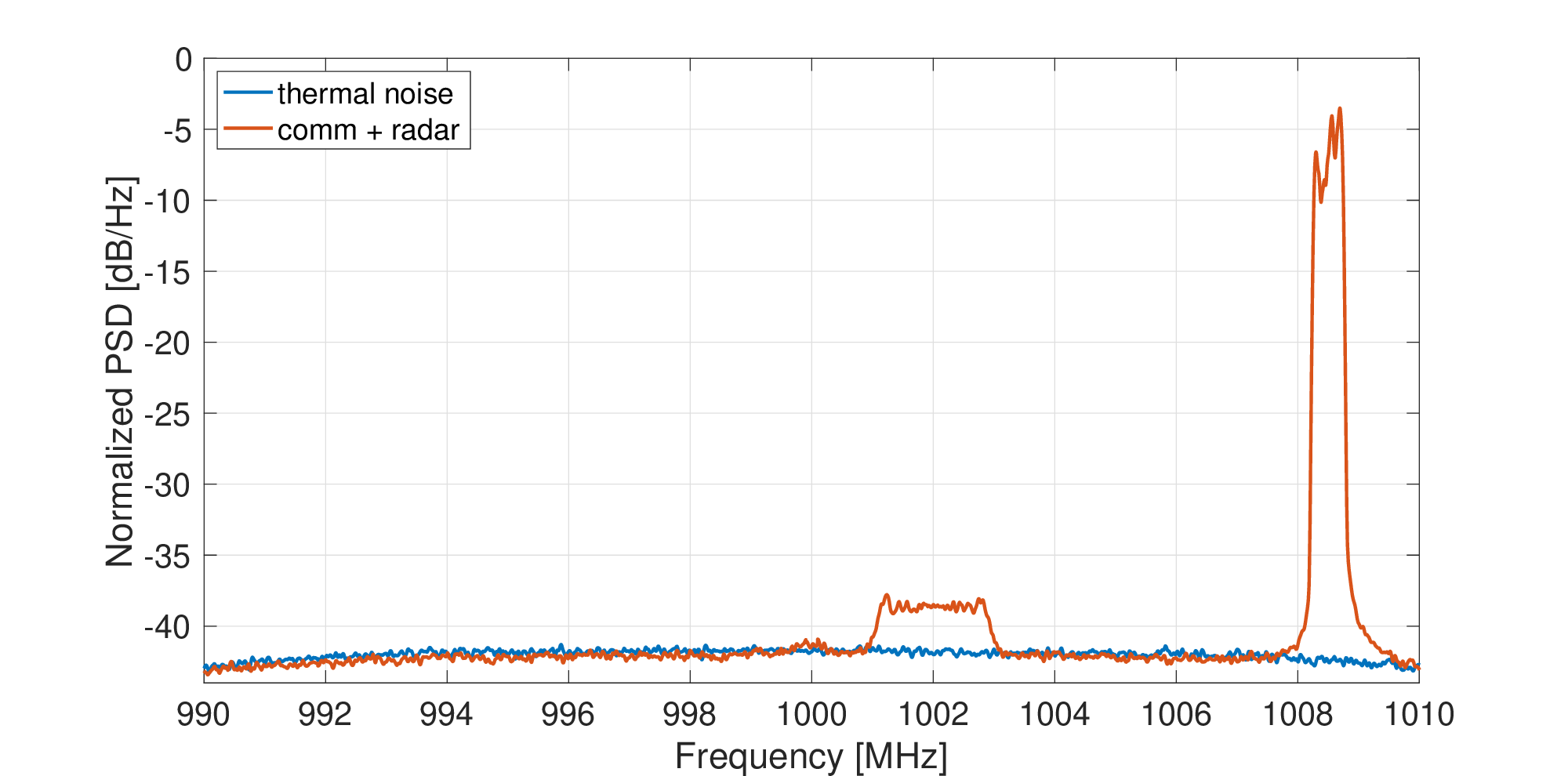}} 
	\hfill \subfloat[]{%
	\includegraphics[trim={35 0 45 10}, clip,width=0.48\linewidth]{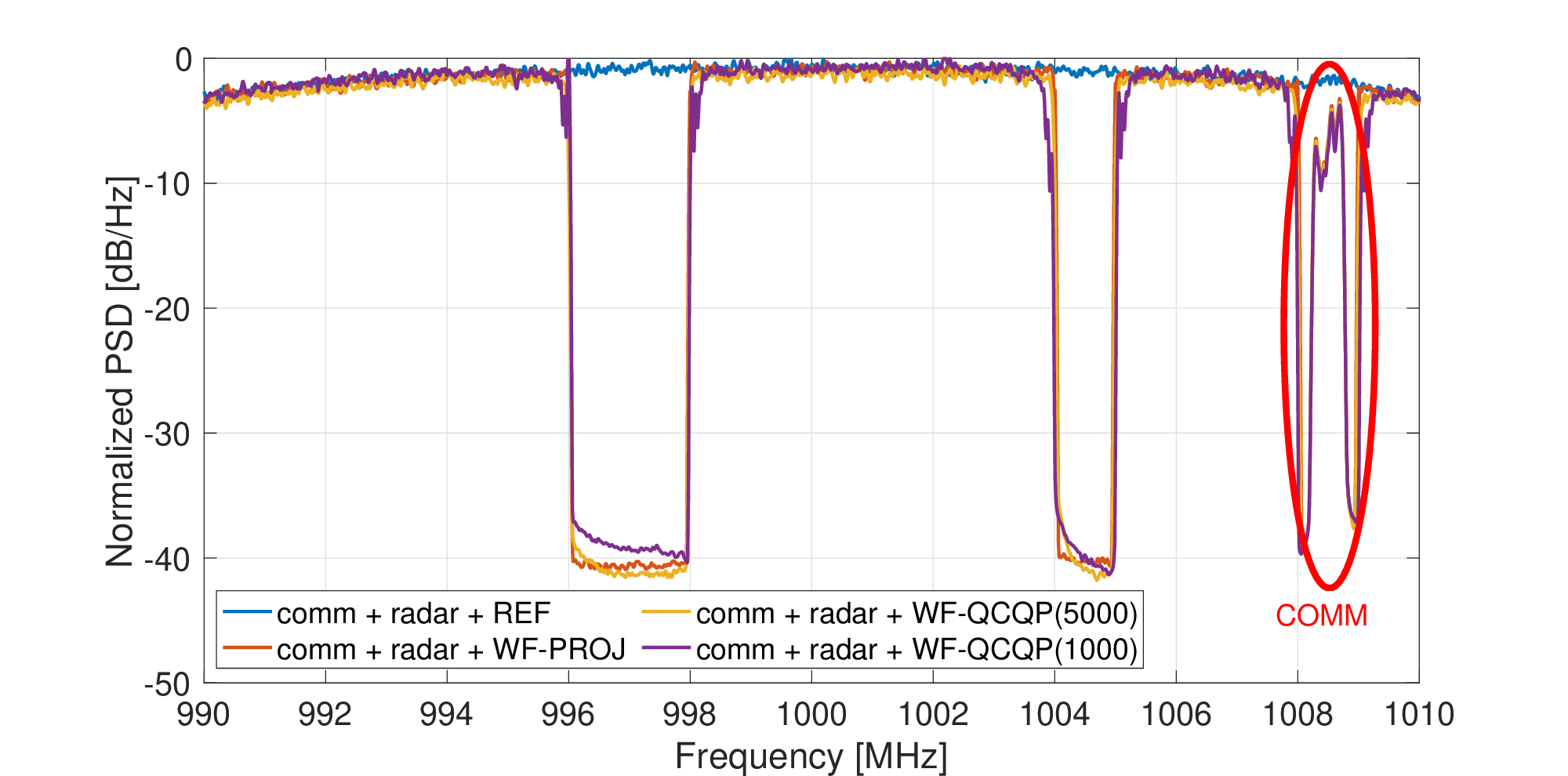}} 
	\caption{PSD estimate of measured data comprising: {(a) thermal noise and communication-plus-radar signal; (b) communication-plus-radar-plus-jamming signal} (reference and tailored waveforms {obtained} via \textbf{Algorithm 1} using block-size either of 1000 or 5000 and \textbf{Algorithm 2}). The reported PSD are estimated using the Welch method (considering segments of 1000 samples with 50$\%$ overlap and weighted with an Blackman-Harris window) and are normalized to the {mean} value of the PSD pertaining to the REF signal.}
	\label{fig:exp_PSD} 
\end{figure*}

\begin{figure*}[t] 
	\centering
	\subfloat[]{%
		\includegraphics[width=0.195\linewidth]{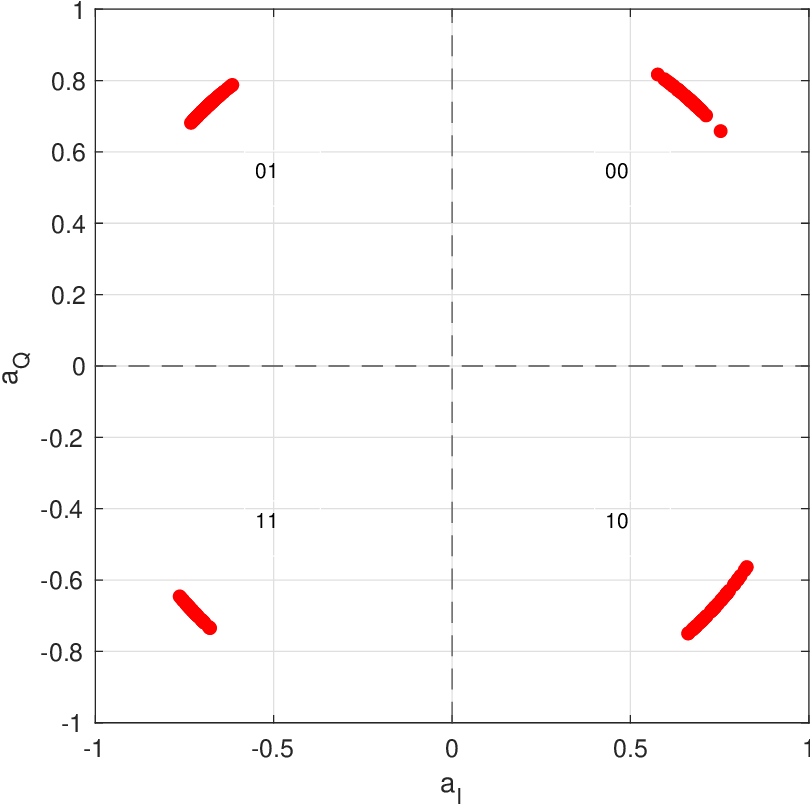}} \hfil
	\subfloat[]{%
		\includegraphics[width=0.195\linewidth]{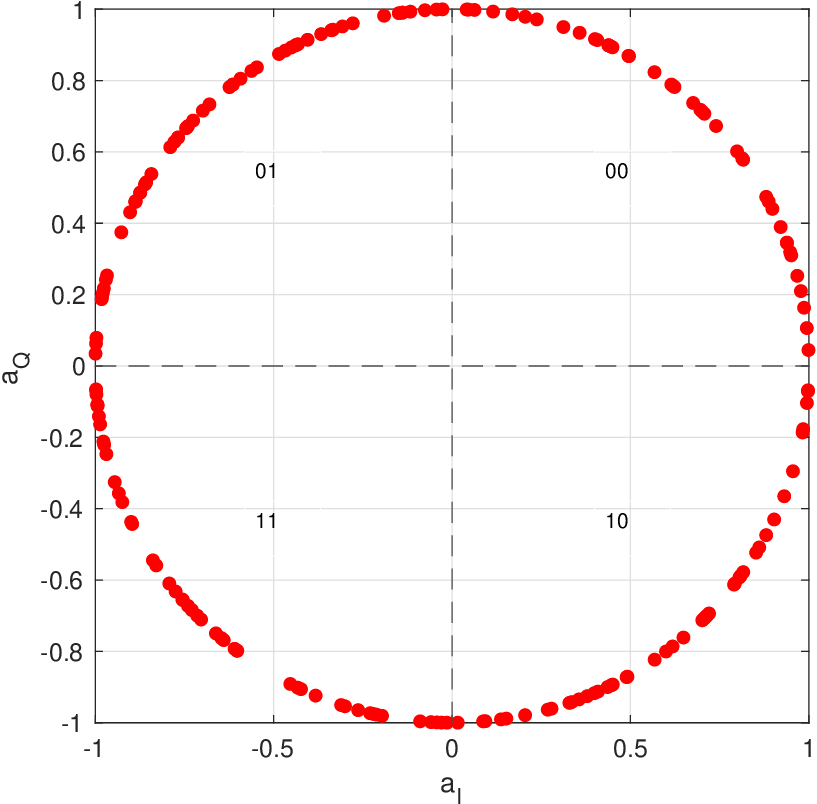}} \hfil
	\subfloat[]{%
		\includegraphics[width=0.195\linewidth]{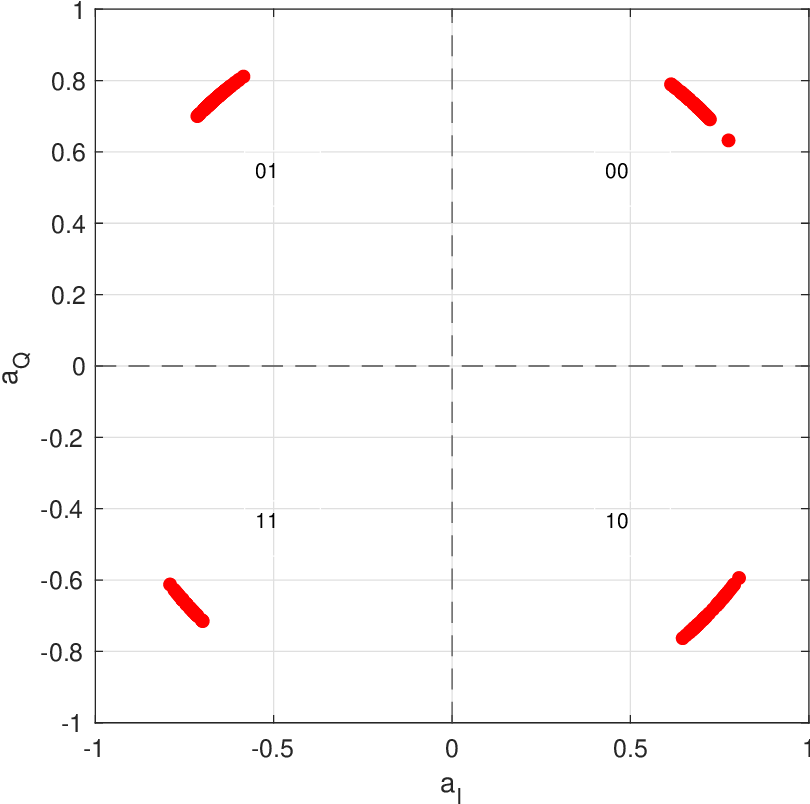}} \hfil
	\subfloat[]{%
		\includegraphics[width=0.195\linewidth]{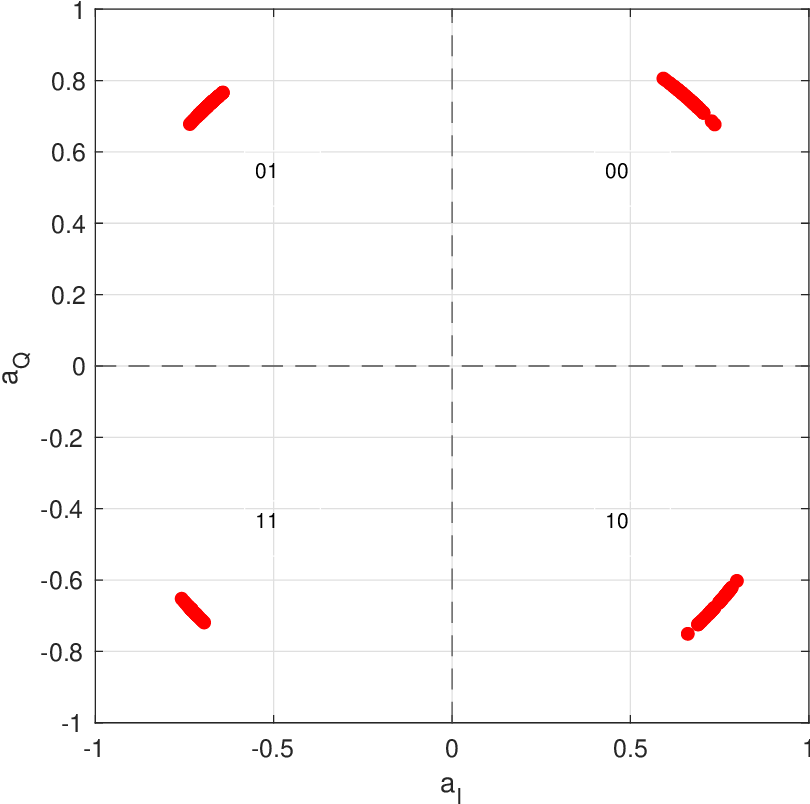}} 	\hfil
	\subfloat[]{%
		\includegraphics[width=0.195\linewidth]{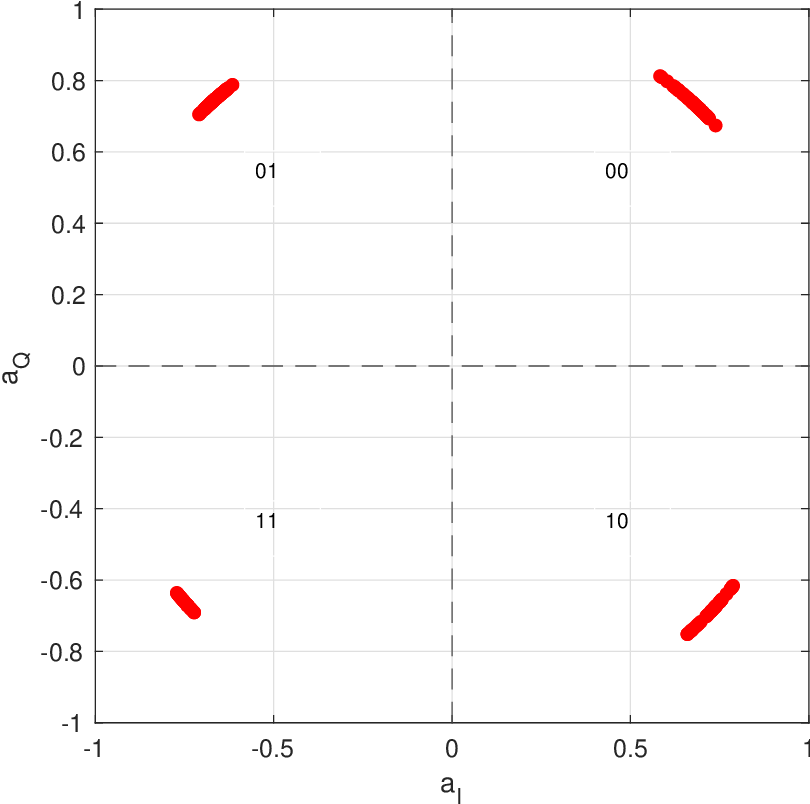}} 
	\caption{Scatterplot of phase transitions between consecutive symbols {for a single communication signal}, evaluated using {data} comprising the superposition of the communication signal with the radar signal and (a) thermal noise; (b) reference waveform; (c) WF-PROJ; (d) WF-QCQP-5000; (e) WF-QCQP-1000.}
	\label{fig:scatter_plot} 
\end{figure*}

In this section, the capability of the proposed algorithms to synthesize spectrally-notched jamming waveforms, ensuring spectral coexistence with friendly frequency-overlaid emitters, is evaluated using measured data. Specifically, the assessment is conducted {employing} a bespoke hardware-in-the-loop testbed {so as} to experimentally verify whether the interference {injected} by the designed jamming signal in the shared frequency bands compromises the {target} radar capabilities  while preserving the communication link of {non-adversarial} communication systems. The analysis requires the following operations:
\begin{itemize}
	\item the simultaneous transmission of:
	\begin{itemize}
		\item a properly {generated} spectrally-notched jamming waveform;
		\item a {frequency-overlaid} communication signal {(friendly system)};
		\item a radar signal {(opponent system);}
	\end{itemize}
	\item the measurement of the {overall} signal within the frequency band of interest;
	\item the evaluation of the error rate in decoding the communication {waveform};
	\item the evaluation of the radar SINR.
\end{itemize}
{Fig.~\ref{fig:test-bed} shows the hardware-in-the-loop testbed used to carry out the experiment,} which is composed of the following devices: 
\begin{itemize}
	\item an ADALM-PLUTO to transmit the communication signal;
	\item an ADALM-PLUTO to transmit the jamming waveform, i.e., {either} a barrage noise or a tailored waveform synthesized {through one of the proposed jamming scheme};
	\item an ADALM-PLUTO to transmit the radar probing signal, i.e., a complex chirp with a bandwidth of 2 MHz;
	\item a Mini-Circuits ZN4PD-4R722+ {combiner} to {add} the jamming, the radar, and {the} communication signals;
	\item an ADALM-PLUTO to collect the {overall} signal for spectral assessment, communication decoding, and radar SINR evaluation;
	\item a 10 dB attenuator to {prevent damage to the} receiver.
\end{itemize}

The ADALM-PLUTO is a SDR device from Analog Devices Inc.~\cite{wyglinski2018software}. It is capable of transmitting in the frequency range $0.3-3.8$ GHz using a channel bandwidth from 200 kHz to 20 MHz. As a receiver, it can acquire signals in the frequency range from 70 MHz to 6 GHz, employing an instantaneous bandwidth up to 20 MHz, and providing the complex envelope of the received signal. Detailed specifications of the ADALM-PLUTO SDR architecture and its components can be found in~\cite{wyglinski2018software}.

In the conducted experiments, the same non-coherent jamming waveforms synthesized and analyzed in Section~\ref{section:numerical_sim} are actually transmitted. {In this respect}, it is supposed that the ESM system has already gathered awareness of the EM environment (via a {sensing phase as contextualized in the PAC}) and identified three friendly RF systems (as detailed in Section~\ref{section:numerical_sim}), including a communication emitter operating within the frequency interval $1008.25 - 1008.75$ MHz.

As to the communication system, a data signal modulated according to a $\pi/4$ differential quadrature phase-shift keying (DQPSK)~{\cite{molisch2012wireless}} is transmitted with a carrier of $1008.5$ MHz, and occupying an RF bandwidth of 500 kHz. The $\pi/4$-DQPSK is a digital modulation technique wherein the data bit stream is mapped into a sequence of symbols, each carrying two bits of information. With the classic phase-shift keying (PSK) scheme, each symbol is modulated via a specific {and fixed} phase value, whereas the $\pi/4$-DQPSK {applies a specific} phase transition to the previous modulation {to carry the current information symbol}. The {mentioned transitions} can take values of $\pi/4$, $3\pi/4$, $-\pi/4$, or $3\pi/4$ to encode the bit pairs $00$, $01$, $10$, $11$, respectively.

Regarding the radar emitter, a standard complex chirp is considered as probing signal, with a bandwidth of 2 MHz and a carrier frequency of 1002 MHz, i.e., occupying the spectral range $1001 - 1003$ MHz.

For the experiment, one SDR is devoted to the transmission of the non-coherent jammer signal, i.e, either the reference signal (barrage noise) or one of the waveforms obtained using \textbf{Algorithm 1} (with block-size either of length $\bar{N}=1000$ or $\bar{N}=5000$ and $W=\bar{N}/2$) or \textbf{Algorithm 2}, whereas the other two are dedicated to {the transmission of} the communication and the radar signal, respectively. The fourth SDR collects the data.

That said, the conducted experiments refer to a operative bandwidth of $20$ MHz centered at 1 GHz. {Moreover, the signals are collected over a time interval of 30 ms with the ADALM-PLUTO sampling frequency set to an high sampling frequency (in order to minimize the effect of digital filtering in the SDR receive chain), i.e., 60 MHz, resulting in $L = 1800000$ complex samples per interval.} Notably, all the acquisitions are performed without employing automatic gain control. {Otherwise stated,} a manual regulation of the receive gain is performed to maximize the dynamic range of the receiver while preventing clipping of the collected signals, regardless of the setup {(the same gain has been used over the entire measurement campaign)}.

Fig.~\ref{fig:exp_PSD} illustrates the {PSD estimate} of the measured data, {normalized to the maximum value of the PSD pertaining to the reference barrage waveform (REF). Fig.~\ref{fig:exp_PSD} (a) illustrates the {PSD estimate} of the communication and the radar signal, as well the {absence} of transmitted signals, showing that the thermal noise {level} is in the order of $-42$ dB/Hz.
The PSD estimate of the received signal when both the jamming (either barrage or a spectrally-notched waveform) and the communication-plus-radar signals are {active} is reported in Fig.~\ref{fig:exp_PSD} (b).
Inspection of the plots clearly highlights the presence of the communication signal around $1008.5$ MHz when a spectrally shaped waveform is employed by the jammer. In contrast, when the reference signal is transmitted (blue curve in Fig.~\ref{fig:exp_PSD} (b)), the communication signal is completely buried. Notably, regardless of the transmitted jamming waveform, the {target} radar signal is always masked by the intentional interfering signal. Moreover, the figures do not show particular differences, in terms of PSD, between the synthesis based on \textbf{Algorithm 1} and \textbf{Algorithm 2}. Precisely, with both the approaches, the measured notch level is in the order of 40 dB, which is different from the designed value. Indeed, since the nominal notch depth for the waveforms synthesized with \textbf{Algorithm 1} and \textbf{Algorithm 2} is 60 dB and deeper than 100 dB, respectively, the expected energy level of the corresponding notch falls below the noise floor of $-42$ dB/Hz. As a consequence, without the communication signal, the measured level in the stop bands is primarily due to the thermal noise.
It is also worth noting that the PSD estimates of the transmitted jamming signals exhibit a slight parabolic shape induced by the ADALM-PLUTO ADC processing {digital filter}.

The effectiveness of the proposed {framework} in ensuring spectral coexistence {with the friendly emissions} is further corroborated by examining the average {communication} error rate. Precisely, {for each measured data}, the received {useful} signal is demodulated and decoded. Subsequently, the decoded information bit sequence (composed of 368 bits) is compared with its nominal counterpart, and the empirical error rate is evaluated. {The results, averaged over 50 acquisition interval, are reported in Table~\ref{table:metric}}. The results show that there {are} no communication errors when the jamming signal is absent as well as when the cognitive jammer {is active}, thereby demonstrating {cohabitation} among these systems. However, when the jammer transmits the reference waveform, it is {generally} not possible to decode {correctly} the message, with a resulting error rate of 49.87\%. In this regard, Fig.~\ref{fig:scatter_plot} shows a scatterplot of phase transitions between consecutive symbols {evaluated for a single communication signal} in the presence of the different jamming waveforms. Here, for a given phase transition $\theta_d$, $a_I = \cos(\theta_d)$, and $a_Q = \sin(\theta_d)$. The plots emphasize that the estimated phase transitions are spread around the {unit circle} when the reference waveform (i.e., a barrage noise) is transmitted. {In contrast,} when the received signal is given by the superposition of the communication and the frequency selective non-coherent waveform, the phase transitions are {quite} concentrated around the corners, with resulting scatter-plots very similar to the benchmark case {without} jamming {activity}, reported in Fig.~\ref{fig:scatter_plot} (a).

Finally, as a figure of merit for radar performance, the SINR of the received probing signal is evaluated under the various experimental scenarios and for different numbers of coherent pulses. Precisely, after matched filtering the collected data and coherently integrating the resulting signal over $M$ pulses, the SINR is estimated as
\begin{equation}
	\text{SINR} = \frac{{\max_h |\bm{y}_\text{R}(h)|^2}}{\frac{1}{H} \sum_{h=1}^H |\bm{y}_\text{J}(h)|^2},
\end{equation}
where $\bm{y}_\text{R}$ is the matched-filtered coherently-integrated received radar signal, and $\bm{y}_\text{J}$ is the matched-filtered output of the collected data comprising only thermal noise and jamming waveform, with $H$ the number of samples analyzed by the radar processor.

The results are illustrated in Table~\ref{table:metric}, which highlights that, in the absence of jamming signals, the single-pulse SINR is 42.13 dB, while integrating 20 and 30 pulses increases the SINR to 55.65 dB and 57.10 dB, respectively, due to the corresponding pulse integration gain. Conversely, when the reference barrage waveform (REF) is transmitted, the single-pulse SINR drops drastically to -15.37 dB and remains insufficient for radar applications even with 30 integrated pulses, with {corresponding} values of only -0.66 dB.
Notably, when one of the synthesized cognitive jamming waveforms is transmitted (WF-PROJ, WF-QCQP-5000, or WF-QCQP-1000), similar SINR values are observed compared to the REF case, thereby demonstrating the effectiveness of the devised approach. These results validate the capabilities of the proposed methods to design non-coherent jamming waveforms that not only hinder enemy RF systems effectively but also ensure coexistence with friendly emitters.

\begin{table*}[t]
	\centering
	\caption{Average communication and radar performance metrics (computed over 50 snapshots).}\label{table:metric}
	\begin{tabular}{l|ccccc}
		\hline
		\hline
		\textbf{Metric} & \textbf{Thermal Noise} & \textbf{REF} & \textbf{WF-PROJ} & \textbf{WF-QCQP-5000} & \textbf{WF-QCQP-1000} \\
		\hline
		{Average Error Rate (in \%)}               & 0       & 49.57       & 0       & 0       & 0       \\ \hline
		Radar SINR (1 pulse) dB & 42.13    & -15.37     & -15.42  & -15.35  & -15.40  \\
		Radar SINR (20 pulses) dB & 55.65  & -2.45      & -2.24   & -2.69   & -2.61  \\
		Radar SINR (30 pulses) dB & 57.10 & -0.66      & -0.54    & -0.85    & -1.04  \\
\hline
\hline
\end{tabular}
\end{table*}

\section{CONCLUSION}
	In this paper, two procedures {have been} proposed for {the} cognitive design {of} selective {non-coherent} jamming signals. In both approaches, the aim is to hinder hostile RF systems (e.g., {radars} or communication infrastructures) by strategically reducing their SINR without interfering with friendly emitters operating within the jammer bandwidth. In particular, at the perception stage, the presence of non-adversarial emitters is recognized within the bandwidth of interest, whereas jamming signals with appropriate frequency allocations {are} synthesized and transmitted along the action stage. To this end, two design approaches have been pursued.
	{The former} allows controlling the amount of interfering energy produced in the bandwidth of {the} friendly emitters and promotes some desirable {noisy} waveform characteristics. However, since the original formulation demands solving a rather high computational and space complex QCQP optimization problem, a {computationally affordable} implementation is proposed by partitioning the jamming waveform in several blocks and sequentially optimizing each of them.
	The {latter} devised technique is a computationally efficient approach based on a suitable projection of the noise-like reference signal onto the subspace orthogonal to the steering vectors spanning the stop frequency bands, allowing to only control the position and width of spectral notches.
The effectiveness of the proposed techniques has
	been numerically demonstrated in terms of spectral features
	and autocorrelation behaviours. The effect of digital
	quantization on the spectral characteristics of such waveforms
	has been investigated. The capabilities of the
	devised approaches have been experimentally corroborated
	using an hardware-in-the-loop testbed with SDRs devices. The
	results have underlined that both methods are effective in
	synthesizing non-coherent jamming waveforms with precisely
	located spectral notches, set according to the information
	acquired during the perception phase. Finally, the transmitted
	cognitive waveforms have proved to successfully ensure
	spectral coexistence with active friendly RF sources while simultaneously impairing the radar capability of an adversarial system by significantly reducing its SINR.

	As an important follow on, there is the integration of the signal synthesis algorithm (which is the action part of the cognitive loop) with modern artificial intelligence-electronic warfare techniques~\cite{haigh2021cognitive} to realize the perception stage. \blue{Finally, it could be worth investigating a different design approach that includes an additional term in the objective function accounting for the increase of the out-of-band energy.}

\bibliographystyle{ieeetr}
\bibliography{ref}

\end{document}